\documentclass[sigconf, nonacm]{acmart}

\newcommand\vldbdoi{XX.XX/XXX.XX}
\newcommand\vldbpages{XXX-XXX}
\newcommand\vldbvolume{18}
\newcommand\vldbissue{11}
\newcommand\vldbyear{2025}
\newcommand\vldbauthors{\authors}
\newcommand\vldbtitle{\shorttitle} 
\newcommand\vldbavailabilityurl{https://github.com/benzhaotang/LogLite}
\newcommand\vldbpagestyle{empty}

\usepackage{multirow} 
\usepackage{subfigure}
\usepackage{subcaption}
\usepackage{tikz}
\usepackage{pgfplots}
\usepackage{color}
\usepackage{xcolor}
\usepackage{graphicx}
\usetikzlibrary{patterns}
\usepackage{siunitx}
\usepackage[normalem]{ulem}
\useunder{\uline}{\ul}{}
\usepackage{hyperref}
\usepackage{bm} 
\usepackage{colortbl}
\usepackage{enumitem}

\definecolor{c1}{RGB}{42,99,172} %
\definecolor{c2}{RGB}{255,88,93}
\definecolor{c3}{RGB}{255,181,73}
\definecolor{c4}{RGB}{119,71,64} %
\definecolor{c5}{RGB}{228,123,121} %
\definecolor{c6}{RGB}{208,167,39} %
\definecolor{c7}{RGB}{0,51,153}
\definecolor{c8}{RGB}{56,140,139} 
\definecolor{c9}{RGB}{0,0,0} 
\definecolor{color1}{HTML}{FFF4D8}
\definecolor{color2}{HTML}{FF7163}

\definecolor{c10}{RGB}{102,205,170} 
\definecolor{c11}{RGB}{255,149,0}    
\definecolor{c12}{RGB}{173,216,230}  
\definecolor{c13}{RGB}{255,204,0}    
\definecolor{c14}{RGB}{153,0,153}    
\definecolor{c15}{RGB}{255,105,180}  

\definecolor{c16}{RGB}{75,192,192}    
\definecolor{c17}{RGB}{255,215,0}     
\definecolor{c18}{RGB}{54,162,235}    
\definecolor{c19}{RGB}{255,192,203}   
\definecolor{c20}{RGB}{144,238,144}   
\definecolor{c21}{RGB}{240,128,128}   
\definecolor{c22}{RGB}{30,144,255}    
\definecolor{c23}{RGB}{255,255,224}   
\definecolor{c24}{RGB}{220,20,60}     
\definecolor{c25}{RGB}{255,192,203}   
\definecolor{c26}{RGB}{0,158,115}     
\definecolor{Blood}{HTML}{860309}
\definecolor{Olive}{HTML}{807805}
\definecolor{Topaz}{HTML}{F5C678}
\definecolor{American Yellow}{HTML}{F3AA07}
\definecolor{Giants Orange}{HTML}{FF5C1E}
\definecolor{Persian Plum}{HTML}{6C1D2A}
\definecolor{Pearl Aqua}{HTML}{85D6B2}
\definecolor{Moonstone}{HTML}{3F9EBD}
\definecolor{Blue Jeans}{HTML}{5BBCF0}
\definecolor{St. Patrick's Blue}{HTML}{2B2D7C}
\definecolor{Green}{HTML}{159879}
\definecolor{Green2}{HTML}{31D683}
\definecolor{Purple}{HTML}{C1689C}

\newcommand{\MYLOGNAME}{LogLite}

\pgfdeclarepatternformonly{myDenseLines}{\pgfqpoint{-1pt}{-1pt}}{\pgfqpoint{2pt}{2pt}}{\pgfqpoint{1.8pt}{1.8pt}}%
{
  \pgfsetlinewidth{0.8pt} 
  \pgfpathmoveto{\pgfqpoint{0pt}{0pt}}
  \pgfpathlineto{\pgfqpoint{2pt}{2pt}}
  \pgfusepath{stroke}
}

\pgfdeclarepatternformonly{myDenseLinesRev}
  {\pgfqpoint{-1pt}{-1pt}}
  {\pgfqpoint{2pt}{2pt}}
  {\pgfqpoint{1.8pt}{1.8pt}}
  {
    \pgfsetlinewidth{0.8pt}
    \pgfpathmoveto{\pgfqpoint{0pt}{2pt}}
    \pgfpathlineto{\pgfqpoint{2pt}{0pt}}
    \pgfusepath{stroke}
  }

\pgfdeclarepatternformonly{myCrosshatch}
  {\pgfqpoint{-1pt}{-1pt}}
  {\pgfqpoint{2pt}{2pt}}
  {\pgfqpoint{1.8pt}{1.8pt}}
  {
    \pgfsetlinewidth{0.5pt}
    \pgfpathmoveto{\pgfqpoint{0pt}{0pt}}
    \pgfpathlineto{\pgfqpoint{2pt}{2pt}}
    \pgfusepath{stroke}
    \pgfpathmoveto{\pgfqpoint{0pt}{2pt}}
    \pgfpathlineto{\pgfqpoint{2pt}{0pt}}
    \pgfusepath{stroke}
  }

\pgfdeclarepatternformonly{myHorizontalLines}
  {\pgfqpoint{-1pt}{-1pt}}
  {\pgfqpoint{2pt}{2pt}}
  {\pgfqpoint{1.8pt}{1.8pt}}
  {
    \pgfsetlinewidth{1pt}
    \pgfpathmoveto{\pgfqpoint{0pt}{1pt}}
    \pgfpathlineto{\pgfqpoint{2pt}{1pt}}
    \pgfusepath{stroke}
  }

  \pgfdeclarepatternformonly{myDots}
  {\pgfqpoint{-1pt}{-1pt}}
  {\pgfqpoint{2pt}{2pt}}
  {\pgfqpoint{1.8pt}{1.8pt}}
  {
    \pgfsetlinewidth{0.5pt}
    \pgfpathcircle{\pgfqpoint{1pt}{1pt}}{0.5pt}
    \pgfusepath{fill}
  }

\begin{document}
\title{\MYLOGNAME{}: Lightweight Plug-and-Play Streaming Log Compression}

\author{
  Benzhao Tang$^{\S}$, Shiyu Yang$^{\S^\ast}$,  Zhitao Shen$^{\dagger}$, 
  Wenjie Zhang$^{\diamond}$, Xuemin Lin$^{\ddagger}$, Zhihong Tian$^{\S^\circ}$
}

\affiliation{
    $^\S$Cyberspace Institute of Advanced Technology of Guangzhou University \& Huangpu Research School of Guangzhou University;
    $^\dagger$Ant Group;  $^{\diamond}$University of New South Wales;  $^{\ddagger}$Shanghai Jiao Tong University; $^{\circ}$Guangdong Key Laboratory of Industrial Control System Security
}
\affiliation{
    benzhaotang@outlook.com; syyang@gzhu.edu.cn; zhitao.szt@antgroup.com
  }
\affiliation{
    wenjie.zhang@unsw.edu.au; xuemin.lin@sjtu.edu.cn; tianzhihong@gzhu.edu.cn
}







\begin{abstract}
Log data is a vital resource for capturing system events and states. With the increasing complexity and widespread adoption of modern software systems and IoT devices, the daily volume of log generation has surged to tens of petabytes, leading to significant collection and storage costs. To address this challenge, lossless log compression has emerged as an effective solution, enabling substantial resource savings without compromising log information. In this paper, we first conduct a characterization study on extensive public log datasets and identify four key observations. Building on these insights, we propose \MYLOGNAME{}, a lightweight, plug-and-play, streaming lossless compression algorithm designed to handle both TEXT and JSON logs throughout their life cycle. \MYLOGNAME{ }requires no predefined rules or pre-training and is inherently adaptable to evolving log structures. Our evaluation shows that, compared to state-of-the-art baselines, \MYLOGNAME{ }achieves Pareto optimality in most scenarios, delivering an average improvement of up to 67.8\% in compression ratio and up to 2.7× in compression speed.
\end{abstract}

\maketitle

\renewcommand{\vldbauthors}{Benzhao Tang, Shiyu Yang, Zhitao Shen, Wenjie Zhang, Xuemin Lin, Zhihong Tian}
\pagestyle{\vldbpagestyle}
\begingroup\small\noindent\raggedright\textbf{PVLDB Reference Format:}\\
\vldbauthors. \vldbtitle. PVLDB, \vldbvolume(\vldbissue): \vldbpages, \vldbyear.\\
\href{https://doi.org/\vldbdoi}{doi:\vldbdoi}
\endgroup
\begingroup
\renewcommand\thefootnote{}\footnote{\noindent
$^{\ast}$Shiyu Yang is the corresponding author.\\
This work is licensed under the Creative Commons BY-NC-ND 4.0 International License. Visit \url{https://creativecommons.org/licenses/by-nc-nd/4.0/} to view a copy of this license. For any use beyond those covered by this license, obtain permission by emailing \href{mailto:info@vldb.org}{info@vldb.org}. Copyright is held by the owner/author(s). Publication rights licensed to the VLDB Endowment. \\
\raggedright Proceedings of the VLDB Endowment, Vol. \vldbvolume, No. \vldbissue\ %
ISSN 2150-8097. \\
\href{https://doi.org/\vldbdoi}{doi:\vldbdoi} \\
}\addtocounter{footnote}{-1}\endgroup

\ifdefempty{\vldbavailabilityurl}{}{
\vspace{.3cm}
\begingroup\small\noindent\raggedright\textbf{PVLDB Artifact Availability:}\\
The source code, data, and/or other artifacts have been made available at \url{\vldbavailabilityurl}.
\endgroup
}

\section{Introduction}
\label{sec:intro}
Logs provide detailed records of various events and states occurring during system operation, including but not limited to user actions, system errors and warnings, security-related activities, and more. 
They serve as essential data sources for system troubleshooting and diagnostics~\cite{he2018identifying,du2017deeplog,meng2019loganomaly,zhang2022deeptralog,zhou2019latent,rosenberg2020spectrum,li2022swisslog}, performance monitoring~\cite{agrawal2018log,yao2018log4perf}, security auditing~\cite{fazzinga2018online,oprea2015detection,amar2019mining}, and business improvements~\cite{dumais2014understanding,fan2015framework}. 
As modern software and IoT systems become increasingly large and complex, coupled with the massive growth in user numbers, logs are being generated at an astonishing rate, ranging from hundreds of terabytes (TB) to tens of petabytes (PB) daily. 
For instance, Alibaba's IoT devices transmit hundreds of TB of logs each day~\cite{aliyunIoT}; 
Uber generates over 10 PB of logs during a busy day across all its services~\cite{wang2024muslope}; 
and large real-time messaging applications like WeChat produce 16--20 PB of logs daily~\cite{yu2023logreducer}.
Due to the importance of logs and regulatory requirements, these substantial volumes of log data must often be collected centrally and preserved losslessly for extended periods. 
Typically, logs are retained for at least six months, and in certain important industries (such as finance and healthcare), retention periods can extend to several years~\cite{wei2023loggrep,liu2019logzip,rodrigues2021clp,wei2021logreducer}.

\begin{figure}[tb]
\begin{tabular}[t]{c}
     \includegraphics[width=0.98\columnwidth]{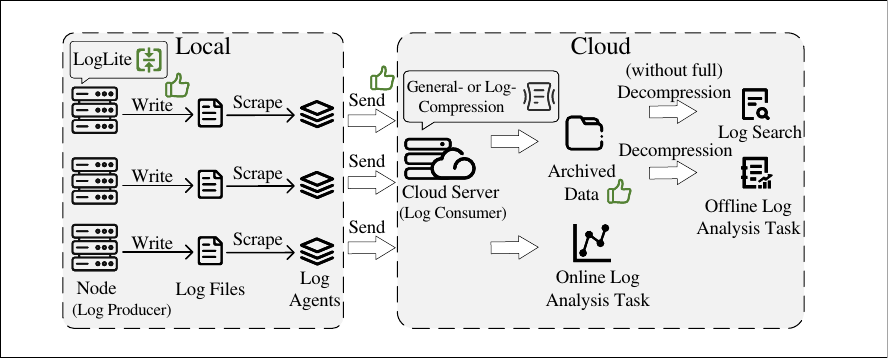}
\end{tabular}
\vspace*{-0.5cm}
\caption{The life cycle of log data. }
\vspace{-1.5em}
\label{fig:introduction}
\end{figure}

To explore the significant cost implications of log collection and storage, we present the complete life cycle of logs from generation to consumption in a distributed cluster, as shown in Figure \ref{fig:introduction}. 
Logs generated independently by multiple nodes responsible for specific services are initially written to local storage. 
Once a certain amount of time elapses or the log file reaches a predefined size, the log agent marks the file as non-writable and scrapes the logs. 
These logs are then sent over the network to dedicated log processing machines or cloud services.
For scenarios with high real-time requirements, these logs are immediately utilized for online log analysis tasks, such as real-time anomaly detection and performance monitoring~\cite{cao2021logstore}. The line-by-line log compression and transmission to the cloud from local nodes support real-time log processing.
For other scenarios, the collected massive log data is compressed using general-purpose or log-purpose compression algorithms and archived to disk. 
When tasks require precise log retrieval, such as root cause analysis for system failures or forensic analysis for regulatory compliance, the archived logs need to be decompressed before log searching. 
Some log-purpose methods~\cite{rodrigues2021clp,wei2023loggrep,wang2024muslope} enable log searches without full decompression. 
However, for offline log analysis tasks, such as resource optimization, business improvement, or security auditing, full decompression is still required. 

Writing logs incurs CPU and I/O overhead, sending logs consumes network bandwidth, and archiving logs requires storage resources~\cite{liu2019logzip,cao2021logstore,yu2023logreducer,ZhengGWYHLG0J23}. 
Given the massive scale of logs, these operations result in significant cost burdens. 
Some log reduction methods~\cite{yu2023logreducer,mastropaolo2022using,rong2023developers,jia2018smartlog} attempt to minimize log generation at the source by selectively generating logs based on necessity, thereby directly reducing the production of non-critical logs. 
However, this approach inevitably sacrifices some log information, and the reduced log volumes often remain within the same order of magnitude, such as decreasing from 19.8 PB to 12.0 PB per day~\cite{yu2023logreducer}. 
As a result, log compression exhibits significant potential to save resources and reduce economic costs. 
Some log systems involve using general-purpose compression algorithms such as LZMA~\cite{lzma} and Zstd~\cite{collet2018zstd}. 
These algorithms do not require any prior knowledge or rules, treating the massive log files as byte streams and replacing repetitive byte sequences with shorter representations. 
However, general-purpose compression methods fail to fully exploit the unique characteristics of logs~\cite{li2024logshrink}.
To address this limitation, some log-purpose compression methods~\cite{liu2019logzip,wei2021logreducer,rodrigues2021clp,wei2023loggrep,wang2024muslope,zhang2023high} have been proposed, which either manually define rules or automatically extract patterns (often referred to as templates) from logs. 
During compression, the pattern of each log is replaced with dictionary entries, requiring only the storage of variables. 
By further applying general-purpose compression, higher compression ratios can be achieved. 
However, after investigating these approaches, we identify the following three challenges:

\textbf{(1) Resource-constrained devices struggle to efficiently compress and transmit logs in real-time.} 
In many scenarios, devices that generate logs have limited resources to store and process them, such as industrial or home IoT devices and nodes focused on their primary tasks~\cite{liu2024adaedge}. 
These devices often prioritize real-time performance monitoring and anomaly detection. 
As a result, logs are expected to be transmitted to cloud services as quickly as possible while consuming minimal resources. 
However, both general-purpose and log-purpose compression methods require the logs to be collected to a sufficient size to achieve effectiveness. 
When logs are stored in small files or have limited samples in resource-constrained devices, these methods struggle to identify redundancies effectively. 
This limitation prevents these methods from delivering effective compression in resource-constrained devices.


\textbf{(2) Most log compression methods require frequent adaptation due to evolving formats, patterns, and structures, limiting their plug-and-play functionality.} 
Logs generated by different devices in diverse deployment environments often contain unique information (e.g., timestamps, IP addresses, and identifiers) and patterns. 
Consequently, most log-purpose methods need to know the predefined templates or be trained to learn the internal structure of the logs, which means they struggle to achieve plug-and-play functionality. 
Moreover, as log information and patterns evolve over time, for example, transitioning from TEXT formats to JSON, methods relying on manually defined rules or patterns extracted through sampling and training become obsolete and require additional resources for redefinition and retraining.


\textbf{(3) Pattern mining or pre-training in existing log compression methods often compromises efficiency, significantly limiting their applicability in time-sensitive scenarios.} Throughout the log processing lifecycle, three key metrics—compression ratio, compression speed, and decompression speed—are critical. However, most log-purpose compression techniques sacrifice both compression and decompression efficiency to accommodate pattern mining and fine-grained processing. This trade-off hinders their effectiveness in high-speed environments, such as real-time log ingestion, where rapid processing is essential.

\begingroup

\renewcommand{\arraystretch}{1} 
\setlength{\tabcolsep}{5pt}  

\begin{table*}[t]
    \centering
    \normalsize
    \caption{Characterization Study of Unstructured(TEXT) and Semi-structured(JSON) Log Datasets.}
    \vspace{-0.5em}
    \label{tab:characterization_study}
    \resizebox{0.9\textwidth}{!}{%
\begin{tabular}{|c|lrr|ccc|ccccc|cc|cc|}
\hline
                       & \multicolumn{3}{c|}{Log Datasets}                                                                                                           & \multicolumn{3}{c|}{Finite Length}                                     & \multicolumn{5}{c|}{Compliance Proportion}     & \multicolumn{2}{c|}{Search Count}                & \multicolumn{2}{c|}{Upper Bound}                    \\ \cline{2-16} 
\multirow{2}{*}{}      & \multicolumn{1}{c|}{\multirow{2}{*}{Name}} & \multicolumn{1}{c|}{\multirow{2}{*}{Size(MB)}} & \multicolumn{1}{c|}{\multirow{2}{*}{\#Lines}} & \multirow{2}{*}{AL} & \multirow{2}{*}{NDL} & \multirow{2}{*}{RDL} & \multicolumn{5}{c|}{PSL   (\%)}  & \multirow{2}{*}{Seq} & \multirow{2}{*}{Rev} & \multirow{2}{*}{MSS (\%)} & \multirow{2}{*}{AN} \\ \cline{8-12}
                       & \multicolumn{1}{c|}{}                      & \multicolumn{1}{c|}{}                          & \multicolumn{1}{c|}{}                         &                     &                      &                      & 32   & 16   & 8    & 4    & 2    &                      &                      &                           &                     \\ \hline
\multirow{16}{*}{\rotatebox{90}{TEXT}} & \multicolumn{1}{l|}{Apache (Apa)}          & \multicolumn{1}{r|}{4.95}                      & 56,482                                        & 90                  & 69                   & 0.1222\%             & 96.4 & 95.8 & 95.1 & 94.4 & 93.5 & 10                   & 2                    & 96.4                      & 9                   \\
                       & \multicolumn{1}{l|}{Linux (Lin)}           & \multicolumn{1}{r|}{2.27}                      & 25,567                                        & 91                  & 208                  & 0.8135\%             & 91.7 & 90.5 & 88.3 & 86.1 & 83.6 & 9                    & 3                    & 92.1                      & 7                   \\
                       & \multicolumn{1}{l|}{Zookeeper (Zoo)}       & \multicolumn{1}{r|}{9.94}                      & 74,380                                        & 139                 & 88                   & 0.1183\%             & 99.7 & 99.7 & 99.6 & 99.5 & 99.2 & 1                    & 1                    & 97.8                      & 7                   \\
                       & \multicolumn{1}{l|}{HealthApp (Hea)}       & \multicolumn{1}{r|}{22.44}                     & 253,395                                       & 92                  & 132                  & 0.0521\%             & 96.3 & 95.6 & 94.8 & 93.3 & 91.2 & 8                    & 2                    & 93.0                      & 7                   \\
                       & \multicolumn{1}{l|}{HPC}                   & \multicolumn{1}{r|}{32.00}                     & 433,490                                       & 76                  & 272                  & 0.0627\%             & 97.5 & 97.2 & 96.8 & 96.3 & 95.3 & 7                    & 2                    & 93.7                      & 8                   \\
                       & \multicolumn{1}{l|}{Android (And)}         & \multicolumn{1}{r|}{183.36}                    & 1,555,005                                     & 123                 & 720                  & 0.0463\%             & 86.5 & 82.2 & 76.9 & 71.1 & 64.9 & 10                   & 7                    & 88.9                      & 10                  \\
                       & \multicolumn{1}{l|}{Hadoop (Had)}          & \multicolumn{1}{r|}{46.35}                     & 395,286                                       & 122                 & 327                  & 0.0827\%             & 98.3 & 97.2 & 95.7 & 93.8 & 87.2 & 3                    & 2                    & 97.7                      & 22                  \\
                       & \multicolumn{1}{l|}{BGL}                   & \multicolumn{1}{r|}{708.76}                    & 4,747,963                                     & 156                 & 235                  & 0.0049\%             & 99.0 & 98.7 & 98.2 & 97.7 & 96.8 & 3                    & 1                    & 93.9                      & 9                   \\
                       & \multicolumn{1}{l|}{Mac}                   & \multicolumn{1}{r|}{16.10}                     & 117,283                                       & 143                 & 490                  & 0.4178\%             & 92.2 & 90.7 & 88.6 & 84.7 & 77.0 & 8                    & 4                    & 92.6                      & 9                   \\
                       & \multicolumn{1}{l|}{OpenStack (Ope)}       & \multicolumn{1}{r|}{38.63}                     & 137,074                                       & 295                 & 131                  & 0.0956\%             & 76.0 & 75.6 & 74.8 & 74.1 & 72.7 & 9                    & 8                    & 90.1                      & 15                  \\
                       & \multicolumn{1}{l|}{Proxifier (Pro)}       & \multicolumn{1}{r|}{2.42}                      & 21,329                                        & 118                 & 104                  & 0.4876\%             & 89.1 & 86.9 & 84.2 & 80.7 & 76.7 & 9                    & 4                    & 93.3                      & 10                  \\
                       & \multicolumn{1}{l|}{Spark (Spa)}           & \multicolumn{1}{r|}{2805.76}                   & 33,236,604                                    & 87                  & 284                  & 0.0009\%             & 84.9 & 83.9 & 83.0 & 82.3 & 81.3 & 7                    & 6                    & 95.3                      & 13                  \\
                       & \multicolumn{1}{l|}{SSH}                   & \multicolumn{1}{r|}{70.02}                     & 655,147                                       & 111                 & 122                  & 0.0186\%             & 98.0 & 97.8 & 97.5 & 96.9 & 96.0 & 3                    & 2                    & 95.6                      & 8                   \\
                       & \multicolumn{1}{l|}{Thunderbird (Thu)}     & \multicolumn{1}{r|}{30320.64}                  & 211,212,192                                   & 153                 & 413                  & 0.0006\%             & 97.7 & 95.9 & 92.7 & 88.8 & 83.5 & 5                    & 3                    & 95.3                      & 11                  \\
                       & \multicolumn{1}{l|}{Windows (Win)}         & \multicolumn{1}{r|}{26716.16}                  & 114,608,388                                   & 243                 & 654                  & 0.0006\%             & 98.3 & 97.0 & 96.5 & 95.1 & 92.9 & 3                    & 2                    & 96.0                      & 16                  \\
                       & \multicolumn{1}{l|}{HDFS}                  & \multicolumn{1}{r|}{1505.28}                   & 11,175,629                                    & 140                 & 135                  & 0.0012\%             & 59.0 & 56.7 & 52.2 & 44.5 & 35.1 & 21                   & 15                   & 87.3                      & 12                  \\ \hline
\multirow{5}{*}{\rotatebox{90}{JSON}}  & \multicolumn{1}{l|}{MongoDB (Mon)}         & \multicolumn{1}{r|}{66355.20}                  & 186,287,600                                   & 395                 & 370                  & 0.0014\%             & 99.9 & 99.9 & 99.9 & 99.9 & 95.7 & 1                    & 1                    & 99.4                      & 14                  \\
                       & \multicolumn{1}{l|}{CockroachDB (Coc)}     & \multicolumn{1}{r|}{10024.96}                  & 24,772,932                                    & 423                 & 2867                 & 0.0116\%             & 94.8 & 93.5 & 91.4 & 88.4 & 84.5 & 1                    & 1                    & 99.4                      & 7                   \\
                       & \multicolumn{1}{l|}{Elasticserch (Ela)}    & \multicolumn{1}{r|}{8171.52}                   & 14,001,234                                    & 611                 & 1194                 & 0.0085\%             & 99.7 & 99.7 & 99.6 & 99.4 & 99.2 & 1                    & 1                    & 97.4                      & 13                  \\
                       & \multicolumn{1}{l|}{Spark}                 & \multicolumn{1}{r|}{2027.52}                   & 1,011,651                                     & 2101                & 1718                 & 0.1698\%             & 97.5 & 96.0 & 93.1 & 89.2 & 84.4 & 3                    & 2                    & 96.4                      & 11                  \\
                       & \multicolumn{1}{l|}{PostgreSQL (Pos)}      & \multicolumn{1}{r|}{392.84}                    & 1,000,000                                     & 411                 & 265                  & 0.0265\%             & 99.5 & 98.9 & 97.7 & 94.6 & 92.6 & 2                    & 2                    & 97.5                      & 13                  \\ \hline
\end{tabular}
    }
\vspace{-1em}
\end{table*}

\endgroup

To effectively and efficiently leverage the characteristics of logs, we conduct a comprehensive characterization study on the relationship between log length and log similarity using 16 public unstructured TEXT datasets~\cite{Loghub} and 5 semi-structured JSON datasets~\cite{wang2024muslope}. 
Through this study, we present four key observations, highlighting a strong correlation between logs of the same length and their likelihood of belonging to the same pattern. 
Based on these insights, we propose a novel and efficient log-purpose compression method, named \MYLOGNAME.
\MYLOGNAME{ }utilizes log length to perform lightweight compression and processes log data line-by-line in a streaming manner. 
For each incoming log, it searches in reverse order within a sliding window that caches logs of the same length to identify similar logs. 
It then applies our custom-designed operations: character-preserving XOR and byte-aligned run-length encoding. 
\MYLOGNAME{ }offers inherent plug-and-play compression for streaming logs, operating automatically without requiring predefined rules, sampling, training, or pattern mining. 
This design enables immediate compression of individual log entries directly at the point of generation for instant transmission. Furthermore, it seamlessly adapts to dynamic changes in log structure without consuming additional resources. 
Consequently, as illustrated in Figure \ref{fig:introduction}, \MYLOGNAME{ }excels in real-time, streaming, and data-agnostic scenarios, such as the transmission of logs from edge nodes or IoT devices to Cloud Servers for online analysis, by significantly reducing resources for log writing and sending while preserving real-time performance. 
This capability stems from key advantages: high compression ratio and speed, zero reliance on prior knowledge, lightweight design, and robust adaptability. 
Furthermore, \MYLOGNAME{ }supports the use of general-purpose compression algorithms during log archiving, similar to other log-purpose compression methods, to achieve additional storage savings.
Our contributions are summarized as follows: 
\begin{itemize}[left=0.07cm, topsep=0em]
    \item We conduct an in-depth characterization study on 21 real-world log datasets from a non-log-parsing perspective and identify four key observations, which reveal for the first time that log length is a highly general and valuable feature for log compression.  

    \item We propose \MYLOGNAME{}, which is, to the best of our knowledge, which is the first lightweight, plug-and-play, streaming, lossless log compression algorithm based on log length to compress both TEXT and JSON logs, offering excellent adaptability. We introduce three core components, including L-Windows, XOR-P and RLE-b/B, that enable \MYLOGNAME{ }to work effectively and efficiently across the entire log life cycle.   

    \item We evaluate \MYLOGNAME{ }against SOTA log-purpose and general-purpose compression baselines on 16 unstructured TEXT and 5 semi-structured JSON log datasets. The results demonstrate that \MYLOGNAME{ }achieves Pareto-optimal performance in most cases.
    
\end{itemize}

The rest of this paper is organized as follows. 
Section \ref{sec:characterizing} describes our extensive characterization study and observations regarding log length. 
Section \ref{sec:overview} introduces the design of \MYLOGNAME{}, a method based on log length, and explains the rationale behind its three core components. 
Section \ref{sec:eva} presents the experimental results of \MYLOGNAME{ }in comparison with SOTA baselines, as well as a case study, ablation studies, and parameter tuning. 
Section \ref{sec:rw} introduces related work on data compression. 
Section \ref{sec:conclusion} concludes the paper.


\section{Characterizing Log Data}
\label{sec:characterizing}

We first introduce the characterization study on 21 widely used log datasets and four observations that motivate the design of \MYLOGNAME{}. 
Previous studies~\cite{wang2017exploiting,li2024logshrink,wang2024muslope} 
conduct characterization analyses on large-scale log data using log parsers. However, our objective is to achieve lightweight streaming compression without parsing logs or relying on prior knowledge. 
Thus, we perform a detailed characterization study of logs from a \textbf{non-log-parsing perspective}.
To the best of our knowledge, this is the first in-depth characterization study of unstructured and semi-structured log datasets that focuses on log length while employing a non-log-parsing approach.



Table~\ref{tab:characterization_study} presents statistics for the log datasets, comprising 16 unstructured TEXT logs from Loghub~\cite{Loghub} and 5 semi-structured JSON logs from \(\mu\)Slope~\cite{wang2024muslope}. To ensure a comprehensive and unbiased study, we incorporate all publicly available datasets from both sources. These datasets are among the most representative and widely adopted in log compression~\cite{li2024logshrink,wang2024muslope,zhang2023high,wei2023loggrep,wei2021logreducer,liu2019logzip} and collectively reflect the diverse characteristics of log data from a broad spectrum of real-world applications. 

First, we define several terms used in the analysis.

\textbf{Log Dataset}: A log dataset \( S = (\log_1, \log_2, \dots, \log_n) \) consists of log entries, where each element \( \log_i \) is a string. The length of the \(n\)-th log entry is denoted by \( len(\log_n) \). 

\textbf{Log Entry}: Each log entry \( \log_n \) is denoted as:
\vspace{-0.5em}
\[
log_n = (c_1, c_2, \dots, c_{len(\log_n)})
\]
where each \( c_i \) represents a character in the \(n\)-th log entry.

\textbf{Similarity Score}: Motivated by the Hamming distance of strings, given two log entries \( \log_n \) and \( \log_m = (c'_1, c'_2, \dots, c'_{len(\log_m)}) \), their similarity score is defined in Equation~\eqref{eq:HSS}:
\begin{equation}
\begin{split}
    Sim(\log_n,\log_m)=\frac{\sum_{i=1}^{\text{len}(\log_n)} \mathbf{1}_{\{c_i = c'_i\}}}{\text{len}(\log_n)}
\end{split}
\label{eq:HSS}
\end{equation}
where \( \text{len}(\log_n) = \text{len}(\log_m) \), and \( \mathbf{1}_{\{c_i = c'_i\}} \) is an indicator function that equals 1 when \( c_i \) is identical to \( c'_i \), and 0 otherwise. 


\textbf{Adjacent Logs}: The adjacent logs of \( \log_n \) are the subset \( (\log_{n-k}, \) \(\dots, \log_{n-1}) \), where \( 1 \leq k  < n \), representing the $k$ preceding logs. Additionally, \( \log_n \) is more adjacent to \( \log_j \) than to \( \log_i \) if and only if \( n-k \leq i < j < n \).


\begin{figure}[tb]
\begin{center}
\begin{tabular}[t]{c}
     \includegraphics[width=0.9\columnwidth]{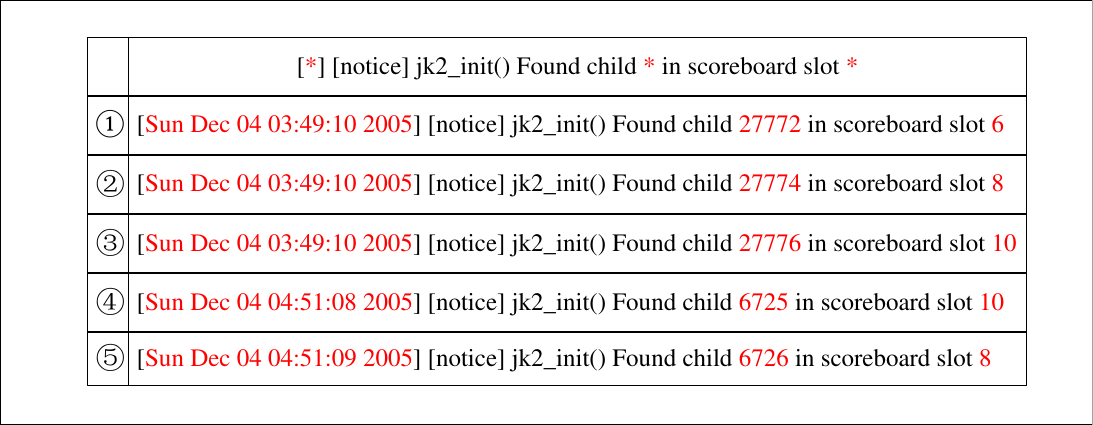}
\end{tabular}
\end{center}
\vspace*{-0.3cm}
\caption{Samples from Apache log.}
\label{fig:Observation_finite}
\vspace{-1em}
\end{figure}


Next, we illustrate the log generation process, using Apache system log generation as an example, which provides insights for our key observations. 
During development, developers write log statements, for example, \texttt{logger.info("jk2\_init() Found child \{child\_id\} in scoreboard slot \{slot\_id\}")}. 
When this log statement is executed, the variables \texttt{child\_id} and \texttt{slot\_id} are filled with real values, a timestamp and other information are added, and it is formatted to produce a log entry such as: \texttt{[Sun Dec 04 03:49:10 2005] [notice] jk2\_init() Found child 27772 in scoreboard slot 6}. Thus, logs typically consist of pre-defined patterns combined with variables populated at runtime. This common structure inherently creates redundancy, which motivates us to derive the following observations for log compression.


\textbf{Observation 1: Log lengths and their variations are practically finite.}
In the above log generation, we observe that each log entry typically consists of an invariant pattern and a small number of variables. 
As shown in Figure~\ref{fig:Observation_finite}, these log entries are generated by a predefined pattern composed of three variables: a timestamp and two integers. 
The length of the pattern is relatively long and fixed, and the length of the variables fluctuates within a relatively small range. 
In practice, since the variables usually represent different values for an identical field, possess consistent data types, and are formatted for readability in log outputs, their length variations tend to remain relatively small. 
For example, in Figure ~\ref{fig:Observation_finite}, the timestamp variable has a fixed length, the second variable typically ranges from 4 to 5 characters, and the third variable typically ranges from 1 to 2 characters. 
The length values for the log entries in this same pattern are 84, 85, and 86. 
Furthermore, for any given log dataset, the number of distinct patterns is finite. 
In most cases, the number of patterns is on the order of a few hundred, with more complex datasets potentially containing over a thousand distinct patterns.

As shown in Finite Length of Table~\ref{tab:characterization_study}, we calculate the \textbf{average length} (AL), \textbf{number of different lengths} (NDL), and \textbf{ratios of different lengths} (RDL) using equations~\eqref{eq:AL},~\eqref{eq:NDL}, and~\eqref{eq:RDL}.

\vspace{-1em}
\begin{equation}
\begin{split}
    AL(S) = \frac{1}{n} \sum_{i=1}^{n} len(\log_i).\\
\end{split}
\label{eq:AL}
\end{equation}
\begin{equation}
\begin{split}
    NDL(S) = \text{count}(\text{unique}(\{len(\log_1), \ldots, len(\log_n)\})).
\end{split}
\label{eq:NDL}
\end{equation}
\begin{equation}
\begin{split}
    RDL(S) = \frac{1}{n} NDL(S).
\end{split}
\label{eq:RDL}
\end{equation}

The average length (AL) of unstructured TEXT logs is generally small, with an average value of 136.
Semi-structured JSON logs, due to their JSON format and the logging strategies of the systems, are larger, with an average length (AL) of 788. 
The average numbers of different lengths (NDL) for TEXT and JSON logs are 274 and 1283, respectively, with average RDL of 0.15\% and 0.04\%, respectively, where each RDL is less than 1\%.
This implies that, in practical applications, both log length and its variation are highly finite and do not increase drastically with growing log volume.




\textbf{Observation 2: Logs with identical lengths have a strong probability of high similarity scores.} 
Since most logs are generated in the above manner, logs belonging to the same pattern have a large number of redundant characters, and these characters are aligned by index because the pattern is fixed and the variation of the variables is very small, which means these redundant characters are suitable for elimination using XOR and run-length coding (e.g., log \textcircled{1} and \textcircled{2} in Figure~\ref{fig:Observation_finite}). 
The more variables that have the same length between two logs, the more redundant characters are aligned. 
Additionally, from Observation 1, we learn that the number of different lengths in a log dataset is finite, and the pattern accounts for the majority of log length.
Consequently, we naturally hypothesize that if two logs have the same length, then they have a high probability of belonging to the same pattern and having variables of the same length, meaning many of their redundant characters are aligned and they have high similarity scores.
To further verify this, we calculate the \textbf{probability of high similarity scores based on length} (PSL), according to formulas ~\eqref{eq:PSL}, which also represents the proportion of logs in dataset $S$ that have a high similarity score exceeding the threshold $\theta$. 
We categorize the $n$ logs from $S$ into $b=NDL(S)$ buckets, where each bucket contains logs of identical length and maintains the relative order of the original file. 
The number of logs in each bucket is denoted as \( (n_1, n_2, \dots, n_b) \).
\begin{equation}
    \begin{split}
        PSL(S) &= \frac{1}{n} \sum_{i=1}^{b} \sum_{j=1}^{n_i} 
        \mathbf{1}\Big\{\exists q \in \{1, 2, \dots, k\} \quad \text{s.t.} \quad j - q > 0 \\
        &\quad \text{and} \quad Sim(\log_j, \log_{j-q}) \geq \theta \Big\}.
    \end{split}
    \label{eq:PSL}
\end{equation}
where the similarity threshold $\bm{\theta}$ is set to a \textbf{default value of 0.85}.

As shown in Compliance Proportion of Table~\ref{tab:characterization_study}, the results show that when the window size $k$ is 32, the PSL exceeds 90\% for most datasets, with some even exceeding 99\%. 
This implies that, given identical length, in a log dataset, most log entries can be matched to a counterpart with a high similarity score, suggesting that length can be leveraged to group and align redundant characters. 
For datasets with low PSL, such as OpenStack and HDFS, our further analysis reveals that their variables are more diverse and change more frequently compared to other datasets. 
After adjusting the similarity threshold \( \theta \) to 0.75, the PSL for both exceed 98.83\%.

We investigate the impact of different window sizes $k$ on the PSL. 
Clearly, as $k$ decreases, the probability of finding similar logs also decreases, since a larger window provides more opportunities to find similar logs. 
Additionally, statistical data show that even with a window size of 2, the average PSL is 82.94\% for TEXT logs and 91.27\% for JSON logs, both of which are notably high.

\begin{figure*}[tb]
\begin{center}
\begin{tabular}[t]{c}
     \includegraphics[width=1.7\columnwidth]{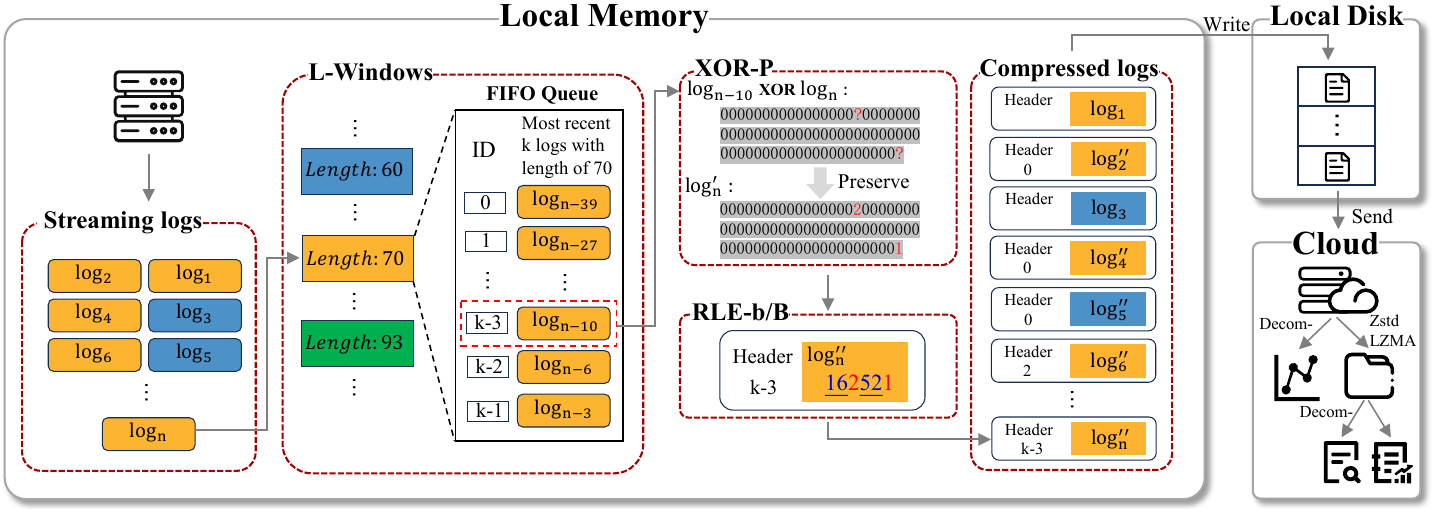}
\end{tabular}
\end{center}
\vspace*{-0.4cm}
\caption{Overview of the \MYLOGNAME{ }framework (Logs of the same length have the same color).}
\vspace*{-1em}
\label{fig:overview}
\end{figure*}

\textbf{Observation 3: More adjacent logs with identical lengths exhibit a stronger probability of high similarity scores.} 
Logs are inherently a type of sequential data, because user and system operations tend to exhibit consistency over short periods. 
For instance, users may repeatedly access the same path or IP within a brief timeframe, many applications and systems maintain a specific state during execution, and multiple threads in concurrent systems may perform the same operations over a short duration. 
Hence, adjacent logs often exhibit considerable consistency in timestamps, user identifiers, system paths, IP addresses, and event types. 


To further quantify, within each bucket of Observation 2, for each log, we search through its $k$ most adjacent logs with identical lengths in sequential and reverse order to find a log whose similarity score with the current log surpasses the similarity threshold $\theta$, continuing until such a log is identified. Here, we set the window size $k = 32$ and use the default value of $\theta$. 
\textbf{Sequential order} refers to searching in the original sequence of the log file, whereas \textbf{reverse order} is the opposite and prioritizes searching from the most adjacent logs first. 
We record the average number of searches needed to find the satisfied logs in both cases, referred to as Seq and Rev, respectively. 
If no matching log entry is found within the $k$ logs, the number of search attempts will be recorded as $k$.

As shown in Search Count of Table~\ref{tab:characterization_study}, on average, the sequential search required 6 attempts to find a similar log, while the reverse search only required 3 attempts, which implies that it is easier to find similar logs when starting the search from more adjacent logs.

\textbf{Observation 4: Logs with identical lengths can achieve significantly high similarity scores.} 
Observation 2 investigates the number of log entries that can be matched to a potentially suboptimal counterpart with a similarity score exceeding the default threshold \( \theta \) (i.e., 85\%). 
To further investigate the upper bound of the average similarity scores a log dataset can achieve, we traverse the \( k = 32 \) most adjacent log entries of the same length to find the optimal counterpart and identify the average \textbf{maximum similarity scores} (MSS) and the theoretical \textbf{average number} (AN) of reverse searches required to find the MSS.

As shown in Upper Bound of Table~\ref{tab:characterization_study}, the MSS for most datasets exceeds 90\%, with an average of 94.7\% across all datasets, which indicates that under best match, a substantial number of characters in a log dataset are redundant, representing the maximum compressible redundancy. 
Additionally, the average number (AN) of reverse searches required is less than 16 in most cases.

\textbf{Discussion.} 
Observation 2 shows that logs of identical length typically follow the same pattern, with the vast majority matching at least one adjacent log within their length bucket. This insight allows \MYLOGNAME{ }to cache and group adjacent logs by length and eliminate redundancy using XOR and RLE (Section \ref{sec:overview}). Given the practical finiteness of log lengths (Observation 1), the memory required to cache $k$ adjacent logs per length remains modest.
Observation 3 reveals that searching in reverse chronological order significantly improves similarity detection efficiency. 
This informs \MYLOGNAME{}'s design to prioritize more adjacent logs, enhancing compression speed. 
While Observation 2 identifies the proportion of suboptimal logs suitable for redundancy elimination, Observation 4 quantifies the maximum of eliminable characters under best match, establishing an upper bound for potential compression gains.

Log length thus serves as an accessible and effective attribute for organizing logs. \MYLOGNAME{ }leverages this characteristic to group, align, and eliminate redundant content across log datasets, enabling both efficient and effective compression.

\section{Methodology}
\label{sec:overview}

\subsection{Preliminaries}
\label{sec:prelim}

The design of the \MYLOGNAME{} compression algorithm is inspired by the bitwise XOR operation on string data. 
Specifically, performing an XOR operation on two identical characters yields a null character (denoted as `$\backslash 0$'), whereas the XOR of two distinct characters results in a non-`$\backslash 0$' character. 
This pivotal characteristic enables a reversible encoding mechanism in which any two original characters, when XORed with the resultant value, can accurately reconstruct the third character. Next, we introduce the detailed solution of \MYLOGNAME{}.

\subsection{Solution Framework}
\label{sec:framework}

In the aforementioned analysis, we observe that log entries with identical length and belonging to the same pattern not only contain a large number of identical characters but also these identical characters are often aligned with each other in terms of positional index. 
The core idea of our solution leverages XOR to convert identical characters into `$\backslash 0$' and then employs run-length encoding (RLE) to achieve efficient and compact compression of these consecutive `$\backslash 0$' characters. 
Based on this concept, we design three components: L-Windows, XOR-P, and RLE-b/B. 

As shown in Figure ~\ref{fig:overview}, a device generates log entries line-by-line, and as these entries are generated, they are immediately compressed line-by-line in local memory using \MYLOGNAME{}, with the compressed logs being output line-by-line. 
Similar to general compression algorithms, \MYLOGNAME{} caches previous logs using L-Windows, while organizing them by length. 
During compression, it identifies similar log entries in the cache to remove redundancy through XOR-P and RLE-b/B. 
Without predefined rules or sampling pre-training, \MYLOGNAME{ }maintains the plug-and-play and adaptability of general compression and leverages the specific characteristics of log data.

In the example of Figure ~\ref{fig:overview}, the length of the orange log entries is 70.
Specifically, to compress the log entry $\log_{n}$ whose length is 70, \MYLOGNAME{ }first locates a corresponding queue whose key is 70 within L-Windows.
The corresponding queue maintains the most adjacent $k$ log entries, each matching the length of $\log_{n}$, from which it selects the entry significantly similar to $\log_{n}$, assumed to be $\log_{n-10}$. 
By performing XOR between $\log_{n-10}$ and $\log_{n}$, it obtains a string having the same length and containing a significant number of consecutive `$\backslash 0$' characters (represented by `0' in the figure). 
XOR-P then preserves the original characters of $\log_{n}$ to replace non-`$\backslash 0$' characters (represented by `$?$' in the figure, where $\log_{n}$ is \textcircled{2} in Figure~\ref{fig:components1}), resulting in $\log^{'}_{n}$ .
The RLE-b/B component then performs compact run-length encoding on $\log^{'}_{n}$, resulting in $\log^{''}_{n}$. 
Additionally, to enable lossless decompression of the log entry, it stores auxiliary information, specifically the header information and the corresponding window ID. 
After compressing $\log_{n}$, the log entries in queue of L-Windows are updated: $\log_{n-39}$ is dequeued from the head, and $\log_{n}$ is inserted at the tail.

Once compressed in memory, the logs are written to local disk files in a streaming manner. 
After a predetermined time interval or when the file size reaches a specified threshold, the file is sent to the cloud for online analysis and archiving. 
As the logs are compressed in local memory, both writing resources and sending network bandwidth are significantly reduced. 
In high-real-time-demand scenarios, individual log entries can be directly sent to the cloud after compression, eliminating time spent writing to disk and waiting for the file to reach the sending condition.

\subsection{Length-Windows}
\label{sec:L-Windows}


Length-Windows (L-Windows) is a hash table with integer keys representing log lengths and values as queues of log entries. 
Each element in the queue is a single log entry. The queue caches up to $k$ log entries whose length corresponds to the key and follows a first-in-first-out (FIFO) policy.


The main idea of L-Windows is to partition log entries that have been compressed into separate queues based on their length and cache them in memory. 
When a new log entry needs to be compressed, it only needs to find a similar log entry among the $k$ entries in the queue whose key is equal to the length of the new log entry, which facilitates subsequent XOR operations to enhance the number of consecutive `$\backslash 0$' characters obtained.
If there is no key equal to the length of the new log entry that needs to be compressed, a new queue is created, and after the new log entry enters the queue, it is directly output without being compressed by XOR-P or RLE-b/B (e.g., $\log_{1}$ and $\log_{3}$ in Figure ~\ref{fig:overview}).

Based on Observation 1, the variation in log length from the same device or service is practically finite, particularly relative to the total size of the logs. 
Therefore, caching log entries into different queues based on their length is highly efficient, and the memory occupied by L-Windows is minimal. 
The overhead of obtaining the length of log entries and the enqueueing and dequeueing times are negligible. 
Theoretically, with $k$ = 8, for a TEXT log dataset, L-Windows requires on average less than 0.3 MB of memory to cache log entries. 
Based on Observations 2 and 3, L-Windows' FIFO strategy ensures that adjacent log entries are cached, and by adjusting the window size $k$, the diversity of cached log entries can be controlled to adapt to the characteristics of different log sources.


\subsection{XOR-Preserve}
\label{sec:XOR-P}

XOR-Preserve (XOR-P) consists of the following two phases: XOR-based Similarity Search and Preserving Original Characters.

\textbf{XOR-based Similarity Search.} To identify a log entry similar to the current log entry being compressed among the $k$ log entries of the same length in the queue of L-Windows, XOR-P performs XOR operations with the log entry at the tail of the queue. 
Then, it calculates the proportion of `$\backslash 0$' characters in the XOR result, which is the similarity score of the two log entries. 
If this proportion exceeds a preset similarity threshold \( \theta \), the log entry is considered similar and selected; otherwise, it performs XOR operations with the next log in the queue from the tail to the head (i.e., in reverse order), until the threshold is reached. 
When the threshold is not even met, the log entry with the highest proportion of `$\backslash 0$' characters is deemed the most similar entry and is selected.

The design of XOR-based Similarity Search is based on the following two rationales: 
(1) Compared to other string similarity calculation methods, such as edit distance, XOR-based similarity score does not require expensive string matching. 
XOR operates at the bit level, is highly efficient on the CPU and is well-suited for SIMD, making it exceptionally fast. 
Additionally, an advantage is that the XOR result of the current compressed log entry and the selected similar log entry is obtained in the XOR-based Similarity Search phases, which can be cached to save the XOR operation in the Preserving Original Characters phase. 
(2) Based on Observation 3, adjacent log entries are often more likely to be similar. Therefore, starting the XOR operation from the most adjacent log entries (i.e., log entries at the tail) rather than from the earliest ones (i.e. log entries at the head) is more likely to meet the similarity threshold.

\begin{figure}[t]
  \centering

  \subfigure[Main Idea of Xor-P]{
    \centering
    \includegraphics[scale=0.42]{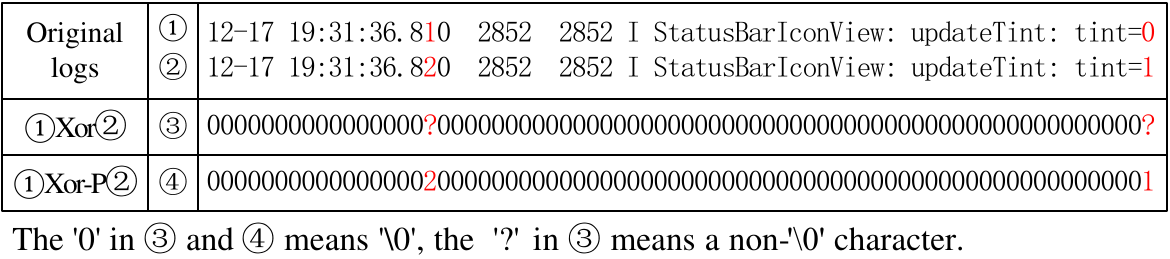}
    \label{fig:components1}
  }\\[3pt] 

  \subfigure[Main Idea of RLE-b/B]{
    \centering
    \includegraphics[scale=0.42]{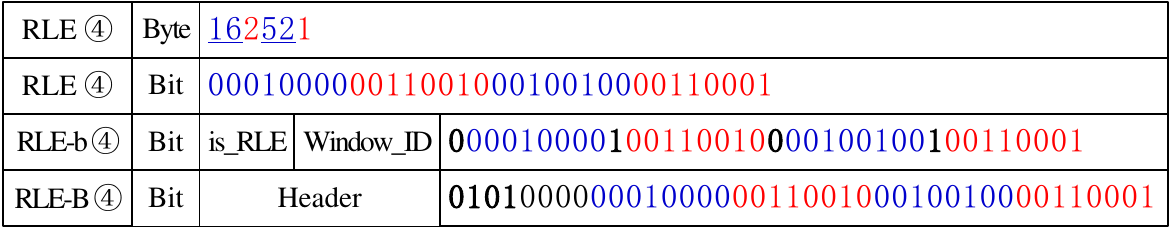}
    \label{fig:components2}
  }\\[3pt] 

  \subfigure[RLE-B Default Format]{
    \centering
    \includegraphics[scale=0.42]{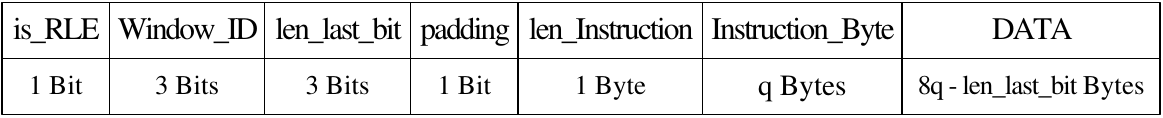}
    \label{fig:components3}
  }
\vspace{-5mm}
  \caption{Components of the \MYLOGNAME{ }framework.}
  \label{fig:component}
  \vspace{-3em}
\end{figure}

\textbf{Preserving Original Characters.} The current log entry being compressed is XOR-ed with the selected similar log entry to obtain an XOR result that contains many consecutive `$\backslash 0$' and some scattered non-`$\backslash 0$' characters. 
As shown in Figure~\ref{fig:components1}, \textcircled{2} represents the current log entry being compressed and \textcircled{1} denotes the selected log entry from the corresponding queue.
\textcircled{2} is XOR-ed with \textcircled{1}, resulting in the XOR result \textcircled{3}. 
For the result, XOR-P then retains all `$\backslash 0$' characters and replaces all non-`$\backslash 0$' characters with the characters at the corresponding indices of the current log entry, so the XOR-P result preserves some original characters. 
As shown in Figure~\ref{fig:components1}, XOR-P transforms \textcircled{3} into \textcircled{4}. 

The design of Preserving Original Characters is based on the following two rationales: 
(1) Preserving Original Characters allows decompression without the need for XOR calculations, thus enhancing decompression speed. 
As shown in Figure ~\ref{fig:components1}, when compressing entry \textcircled{2}, we obtain \textcircled{3} from \textcircled{1} XOR \textcircled{2}. 
To decompress \textcircled{2}, the calculation \textcircled{2} = \textcircled{1} XOR \textcircled{3} must be performed once. 
However, for log entry \textcircled{4}, which preserves the original characters, decompression only requires replacing `$\backslash 0$' with the corresponding characters from \textcircled{1}, eliminating the need for XOR computation.
(2) Compared to storing XOR results, Preserving Original Characters uses the same storage space but preserves more of the original characters of the log entry. 
Since these original characters are natural language, their distribution is more concentrated than that of characters obtained by XOR operations, which can improve compression ratios in subsequent general-purpose compression algorithms.

\vspace{-2em}
\subsection{RLE-bit/Byte}
\label{sec:RLE-b/B}
After applying XOR-P, we obtain the XOR result \textcircled{4} as shown in Figure~\ref{fig:components1}. 
As the RLE \textcircled{4} in Figure~\ref{fig:components2}, the idea of Run-Length Encoding (RLE) is to record only the count of consecutive `$\backslash 0$' characters, while non-`$\backslash 0$' characters (i.e., the original characters from the current log) are fully retained. 
In Figure ~\ref{fig:components2}, it shows both the byte and bit representations of XOR-P result \textcircled{4} after RLE. 
In the RLE \textcircled{4} of byte, the blue 16 and 52 are treated as integers and thereby require only one byte (8 bits) each for storage, while the red `2' and `1'  are treated as characters and also require one byte (8 bits) each. 
Consequently, a log entry originally 70 bytes in size can be losslessly stored using only 4 bytes along with minimal additional metadata, thereby achieving a high compression ratio.

After being compressed and transmitted to cloud server for log processing, logs may need to be parsed in real-time for anomaly detection and monitoring in certain applications, while in others, archiving collected logs suffices. 
Therefore, we design RLE-bit (RLE-b) and RLE-Byte (RLE-B), respectively. 
RLE-b outputs a bit stream, providing high compression ratios and fast processing, making it ideal for real-time log anomaly detection and monitoring. 
However, it is not compatible with further compression by general-purpose algorithms. 
In contrast, RLE-B outputs a byte stream, achieves competitive compression ratio and speed and supports general compression algorithms at the byte level for further compression, making it ideal for log archiving scenarios.

As shown in Figure~\ref{fig:components2}, in the example of RLE-b \textcircled{4}, RLE-b only needs to record \texttt{is\_RLE}, \texttt{Window\_ID} and the streaming indicator bits (i.e., the bits are bolded) as additional metadata. 
These indicator bits specify whether the following 8 bits represent the count of consecutive `$\backslash 0$' characters or an original character from the log. 
This streaming design enables RLE-b to achieve high speeds.



As shown in Figure~\ref{fig:components2}, in the example of RLE-B \textcircled{4}, RLE-B aligns data into a byte stream by grouping the indicator bits into bytes (i.e., \texttt{Instruction\_Byte}). 
Since there are fewer than 8 bits, 4 zero bits are added as padding.
The header contains supplementary details to facilitate lossless decompression. 
The standard RLE-B format is depicted in Figure~\ref{fig:components3}. Here, \texttt{is\_RLE} specifies if run-length encoding is utilized. \texttt{Window\_ID} identifies the position within the L-Windows queue (with $k$ typically set to 8). \texttt{len\_last\_bit} indicates the count of meaningful bits in the final byte of \texttt{Instruction\_Byte}, enabling the decompression process to disregard padding zero bits. 
\texttt{Padding} consists of filler bits that ensure the initial four fields are byte-aligned. \texttt{len\_Instruction} captures the byte length of \texttt{Instruction\_Byte} (denoted as $q$), which inherently establishes that the byte length of \texttt{DATA} is $(8q - \texttt{len\_last\_bit})$, thereby allowing decompression to distinguish between \texttt{Instruction\_Byte} and \texttt{DATA} segments.

\subsection{Decompression}
\label{sec:Decompression}

\MYLOGNAME{ }performs line-by-line streaming decompression, preserving the original log sequence order. 
Similar to most general compression, \MYLOGNAME{ }decompresses a log entry based on previously decompressed log entries.

To decompress the compressed logs shown in Figure ~\ref{fig:overview}, since $\log_{1}$ is the first log with a length of 70 that has not undergone XOR-P and RLE-b/B, \MYLOGNAME{ }creates a new queue in L-Windows according to the length of $\log_{1}$, directly outputs $\log_{1}$ and caches it in the new queue. 
$\log^{''}_{2}$ is decoded using run-length decoding to obtain $\log^{'}_{2}$. 
Then, using $Window\_ID$ (i.e., 0) and the length of $\log^{'}_{2}$ (i.e., 70), \MYLOGNAME{ }finds $\log_{1}$ in the corresponding queue. 
The `$\backslash 0$' characters in $\log^{'}_{2}$ are replaced by the characters at the same indices in $\log_{1}$, reconstructing the original log $\log_{2}$. $\log_{2}$ is output and also cached in the queue of length 70. 
$\log_{3}$ is a log that has not undergone XOR-P and RLE-b/B, so it is directly output and cached in the new queue of length 60. 
The remaining logs repeat the above process to achieve efficient streaming decompression. 


\subsection{Complexity}
\label{sec:Complexity}
Given \( n \) log entries with an average length of \( AL \), and the number of different lengths is \( NDL \). During the compression, it is necessary to maintain an L-Windows, which can have a maximum of \( NDL \) queues, each queue containing a maximum of \( k \) logs having an average length of \( AL \), where k is a constant. 
Therefore, the space complexity is \( O(NDL \times k \times AL) \).


When compressing a log, at most \( k \) logs with an average length of \( AL \) need to be XORed, resulting in a time complexity of \( O(k \times AL) \). 
The time complexity of the encoding of an RLE is \( O(AL) \). 
The time complexity for compressing a log is \(  O( (k+1) \times AL) \).
Thus, the time complexity for compressing \( n \) logs is \( O( n \times (k+1) \times AL) \).



In the worst-case scenario, where all log lengths are different, i.e., \( NDL = n \), and \( AL = \frac{n(n+1)}{2n} = O(n) \), the space complexity becomes \( O(kn^2) \), and the time complexity becomes \( O(kn^2) \). 
However, the characterization study in Table 1 indicates that \( NDL \) and \( AL \) are much smaller than $n$, approximated to constants.
Therefore, in practical applications, the space complexity is very small, and the time complexity is equivalent to \( O(n) \). 
Decompression is the inverse process of compression and shares the same complexity.



\section{Experimental Evaluation}
\label{sec:eva}

To validate our observations and the effectiveness of the proposed \MYLOGNAME{ }solution, we conduct comprehensive experiments on 21 public and widely used log datasets, comparing it against SOTA general-purpose and log-specific compression baselines. Specifically, we focus on the following aspects: 
(1) The compression ratio, compression speed, and decompression speed of \MYLOGNAME{ }for both line-by-line log compression and file compression. 
(2) The throughput of \MYLOGNAME{ }in real-time compression scenarios.
(3) The contribution of each component to the effectiveness of \MYLOGNAME{}.  
(4) The impact of different configurations on \MYLOGNAME{}'s performance. 

\subsection{Experimental Setting}
\subsubsection{\textbf{Datasets.}} 
We use two log datasets, detailed in Table \ref{tab:characterization_study}.  

(1) Unstructured log datasets: 
We include all 16 public unstructured TEXT log datasets from Loghub~\cite{Loghub}. 
HDFS, Had, Spa, Zoo, and Ope are from Distributed Systems (DS); BGL, HPC, and Thu are from Super Computers (SC); Win, Lin, and Mac from Operating Systems (OS); And and Hea are from Mobile Systems (MS); Apa and SSH are from Server Applications (SA); and Pro is from Standalone Software (SS). 
The total size exceeds 60 GB.

(2) Semi-structured log datasets: 
We utilize all 5 public semi-structured JSON log datasets from \(\mu\)Slope~\cite{wang2024muslope}, which are generated by running the HiBench~\cite{huang2010hibench} and YCSB~\cite{ycsb} benchmarks. 
The total size exceeds 80 GB.

\subsubsection{\textbf{Environments.}} 
Our code is implemented in C++ and compiled with g++ 9.4.0 using the -Ofast optimization flag. 
All experiments are conducted on a machine equipped with an Intel(R) Xeon(R) Platinum 8336C CPU @ 2.30GHz, 125GB of main memory, and running Linux kernel version 5.15.0-124-generic.

\begingroup
\renewcommand{\arraystretch}{1.1} 
\setlength{\tabcolsep}{1.5pt}  

\begin{table*}[t]
    \centering
    \small
    \caption{Line-by-line compression performance.(Speed is measured in MB/s)}
    \vspace{-0.5em}
    \label{tab:line_compression}
    \resizebox{\textwidth}{!}{%
\begin{tabular}{|ccc|cccccccccccccccc|ccccc|}
\hline
\multicolumn{3}{|c|}{\multirow{2}{*}{Datasets}}                                                                                                                                                                     & \multicolumn{16}{c|}{TEXT}                                                                                                                                                                                                                                                                                                                                                                    & \multicolumn{5}{c|}{JSON}                                                                                             \\ \cline{4-24} 
\multicolumn{3}{|c|}{}                                                                                                                                                                                              & Apa                   & Lin                   & Zoo                   & Hea                   & HPC                   & And                   & Had                   & BGL                   & Mac                   & Ope                   & Pro                   & Spa                   & SSH                   & Thu                   & Win                   & HDFS                  & Mon                   & Coc                   & Ela                   & Spark                 & Pos                   \\ \hline
\multicolumn{1}{|c|}{\multirow{7}{*}{\rotatebox{90}{Compression Ratio}}}   & \multicolumn{1}{c|}{\multirow{4}{*}{\rotatebox{90}{Specific}}} & PBC                             & 0.185                 & 0.240                 & 0.118                 & 0.284                 & 0.298                 & 0.355                 & 0.156                 & 0.331                 & 0.277                 & 0.190                 & 0.174                 & 0.284                 & 0.193                 & 0.711                 & 0.219                 & 0.320                 & 0.607                 & 0.353                 & 0.103                 & 0.565                 & 0.083                 \\
\multicolumn{1}{|c|}{}                                                                        & \multicolumn{1}{c|}{}                                                             & PBC-F                           & 0.106                 & 0.159                 & 0.075                 & 0.174                 & 0.201                 & 0.251                 & 0.105                 & 0.168                 & 0.197                 & {\ul \textbf{0.126}}  & {\ul \textbf{0.113}}  & 0.158                 & 0.119                 & 0.382                 & 0.114                 & {\ul \textbf{0.144}}  & 0.235                 & 0.225                 & 0.080                 & 0.398                 & {\ul \textbf{0.045}}  \\
\multicolumn{1}{|c|}{}                                                                        & \multicolumn{1}{c|}{}                                                             & \MYLOGNAME{}-B & 0.112                 & 0.175                 & 0.071                 & 0.174                 & 0.154                 & 0.250                 & 0.103                 & 0.127                 & 0.160                 & 0.166                 & 0.163                 & 0.105                 & 0.108                 & 0.116                 & 0.105                 & 0.232                 & 0.024                 & 0.119                 & 0.028                 & 0.096                 & 0.072                 \\
\multicolumn{1}{|c|}{}                                                                        & \multicolumn{1}{c|}{}                                                             & \MYLOGNAME{}-b & {\ul \textbf{0.088}}  & {\ul \textbf{0.157}}  & {\ul \textbf{0.056}}  & {\ul \textbf{0.153}}  & {\ul \textbf{0.131}}  & {\ul \textbf{0.234}}  & {\ul \textbf{0.086}}  & {\ul \textbf{0.114}}  & {\ul \textbf{0.147}}  & 0.159                 & 0.144                 & {\ul \textbf{0.081}}  & {\ul \textbf{0.092}}  & {\ul \textbf{0.103}}  & {\ul \textbf{0.098}}  & 0.218                 & {\ul \textbf{0.019}}  & {\ul \textbf{0.112}}  & {\ul \textbf{0.023}}  & {\ul \textbf{0.095}}  & 0.067                 \\ \cline{2-24} 
\multicolumn{1}{|c|}{}                                                                        & \multicolumn{1}{c|}{\multirow{3}{*}{\rotatebox{90}{General}}}  & FSST                            & 0.381                 & \textbf{0.441}        & 0.386                 & \textbf{0.414}        & 0.418                 & 0.609                 & 0.425                 & \textbf{0.346}        & 0.599                 & 0.437                 & 0.313                 & 0.479                 & 0.322                 & 0.467                 & 0.395                 & \textbf{0.326}        & 0.457                 & 0.792                 & 0.760                 & 0.638                 & 0.320                 \\
\multicolumn{1}{|c|}{}                                                                        & \multicolumn{1}{c|}{}                                                             & LZ4-d                           & \textbf{0.350}        & 0.444                 & \textbf{0.250}        & 0.418                 & \textbf{0.353}        & \textbf{0.569}        & \textbf{0.305}        & 0.382                 & \textbf{0.414}        & 0.265                 & \textbf{0.310}        & \textbf{0.358}        & \textbf{0.302}        & 0.413                 & 0.251                 & 0.375                 & \textbf{0.163}        & 0.173                 & 0.241                 & 0.245                 & 0.154                 \\
\multicolumn{1}{|c|}{}                                                                        & \multicolumn{1}{c|}{}                                                             & Zstd-d                          & 0.428                 & 0.491                 & 0.283                 & 0.477                 & 0.405                 & 0.571                 & 0.314                 & 0.365                 & 0.425                 & \textbf{0.236}        & 0.354                 & 0.399                 & 0.366                 & \textbf{0.408}        & \textbf{0.243}        & 0.344                 & 0.169                 & \textbf{0.116}        & \textbf{0.195}        & \textbf{0.179}        & \textbf{0.136}        \\ \hline
\multicolumn{1}{|c|}{\multirow{7}{*}{\rotatebox{90}{Compression Speed}}}   & \multicolumn{1}{c|}{\multirow{4}{*}{\rotatebox{90}{Specific}}} & PBC                             & 53.6                  & 68.5                  & 129.0                 & 60.7                  & 65.3                  & 52.3                  & 101.0                 & 32.9                  & 94.1                  & 181.9                 & 32.8                  & 52.5                  & 57.0                  & 121.3                 & 105.7                 & 97.2                  & 398.1                 & 225.7                 & 433.4                 & 194.1                 & 167.2                 \\
\multicolumn{1}{|c|}{}                                                                        & \multicolumn{1}{c|}{}                                                             & PBC-F                           & 51.5                  & 64.1                  & 122.2                 & 56.2                  & 62.1                  & 48.8                  & 95.0                  & 31.5                  & 117.1                 & 159.5                 & 32.0                  & 49.7                  & 54.4                  & 94.1                  & 99.2                  & 87.9                  & 237.3                 & 202.4                 & 370.5                 & 135.8                 & 158.5                 \\
\multicolumn{1}{|c|}{}                                                                        & \multicolumn{1}{c|}{}                                                             & \MYLOGNAME{}-B & 100.7                 & 77.8                  & 151.0                 & 79.8                  & 83.2                  & 83.7                  & 118.9                 & 135.5                 & 104.3                 & 138.3                 & 93.1                  & 107.1                 & 110.7                 & 135.3                 & 160.4                 & 91.8                  & 338.2                 & 187.5                 & 457.3                 & 316.4                 & 285.6                 \\
\multicolumn{1}{|c|}{}                                                                        & \multicolumn{1}{c|}{}                                                             & \MYLOGNAME{}-b & \textbf{219.1}        & \textbf{132.5}        & \textbf{333.1}        & \textbf{137.3}        & \textbf{145.8}        & \textbf{122.6}        & \textbf{233.3}        & \textbf{203.7}        & \textbf{167.7}        & \textbf{183.6}        & \textbf{148.7}        & \textbf{226.3}        & \textbf{213.8}        & \textbf{217.4}        & \textbf{217.9}        & \textbf{127.2}        & {\ul \textbf{650.1}}  & {\ul \textbf{260.1}}  & {\ul \textbf{737.8}}  & {\ul \textbf{321.7}}  & {\ul \textbf{406.8}}  \\ \cline{2-24} 
\multicolumn{1}{|c|}{}                                                                        & \multicolumn{1}{c|}{\multirow{3}{*}{\rotatebox{90}{General}}}  & FSST                            & {\ul \textbf{432.2}}  & {\ul \textbf{384.2}}  & {\ul \textbf{351.1}}  & {\ul \textbf{329.3}}  & {\ul \textbf{378.1}}  & {\ul \textbf{239.6}}  & {\ul \textbf{268.6}}  & {\ul \textbf{290.1}}  & {\ul \textbf{256.5}}  & {\ul \textbf{275.1}}  & {\ul \textbf{503.6}}  & {\ul \textbf{247.1}}  & {\ul \textbf{457.8}}  & {\ul \textbf{250.4}}  & {\ul \textbf{274.7}}  & {\ul \textbf{316.8}}  & \textbf{261.1}        & \textbf{182.8}        & \textbf{197.8}        & \textbf{200.6}        & \textbf{291.9}        \\
\multicolumn{1}{|c|}{}                                                                        & \multicolumn{1}{c|}{}                                                             & LZ4-d                           & 29.6                  & 33.1                  & 50.2                  & 33.2                  & 35.4                  & 38.0                  & 56.8                  & 46.8                  & 46.2                  & 75.7                  & 33.6                  & 40.5                  & 23.5                  & 45.9                  & 63.9                  & 47.2                  & 99.0                  & 207.2                 & 179.4                 & 177.4                 & 109.6                 \\
\multicolumn{1}{|c|}{}                                                                        & \multicolumn{1}{c|}{}                                                             & Zstd-d                          & 8.4                   & 8.6                   & 12.8                  & 8.5                   & 9.1                   & 10.7                  & 15.6                  & 12.3                  & 13.3                  & 25.3                  & 10.9                  & 10.6                  & 10.3                  & 12.3                  & 17.2                  & 12.6                  & 29.4                  & 112.7                 & 84.7                  & 101.6                 & 34.7                  \\ \hline
\multicolumn{1}{|c|}{\multirow{7}{*}{\rotatebox{90}{Decompression Speed}}} & \multicolumn{1}{c|}{\multirow{4}{*}{\rotatebox{90}{Specific}}} & PBC                             & {\ul \textbf{2580.8}} & {\ul \textbf{2396.2}} & {\ul \textbf{2578.4}} & {\ul \textbf{2006.9}} & {\ul \textbf{1947.3}} & {\ul \textbf{1967.5}} & {\ul \textbf{2900.1}} & {\ul \textbf{1525.1}} & {\ul \textbf{1532.1}} & \textbf{1975.6}       & {\ul \textbf{1949.5}} & {\ul \textbf{2066.0}} & {\ul \textbf{2676.8}} & {\ul \textbf{1094.5}} & \textbf{1211.6}       & {\ul \textbf{2161.8}} & {\ul \textbf{2640.7}} & \textbf{2311.6}       & {\ul \textbf{3405.3}} & {\ul \textbf{4041.5}} & \textbf{2543.9}       \\
\multicolumn{1}{|c|}{}                                                                        & \multicolumn{1}{c|}{}                                                             & PBC-F                           & 702.4                 & 628.8                 & 462.1                 & 236.4                 & 571.3                 & 650.8                 & 1312.1                & 637.1                 & 544.3                 & 1012.9                & 650.0                 & 1540.4                & 783.8                 & 477.1                 & 704.7                 & 838.7                 & 1535.5                & 2350.4                & 2871.7                & 1833.9                & 1430.2                \\
\multicolumn{1}{|c|}{}                                                                        & \multicolumn{1}{c|}{}                                                             & \MYLOGNAME{}-B & 496.3                 & 379.3                 & 640.6                 & 341.8                 & 409.4                 & 321.0                 & 530.6                 & 571.3                 & 405.0                 & 468.3                 & 412.9                 & 522.4                 & 506.1                 & 561.7                 & 732.8                 & 369.9                 & 1307.5                & 613.9                 & 1251.3                & 755.1                 & 877.9                 \\
\multicolumn{1}{|c|}{}                                                                        & \multicolumn{1}{c|}{}                                                             & \MYLOGNAME{}-b & 493.2                 & 371.7                 & 680.6                 & 356.3                 & 391.2                 & 342.6                 & 539.7                 & 586.9                 & 412.1                 & 485.1                 & 413.4                 & 544.7                 & 514.3                 & 581.0                 & 573.0                 & 378.7                 & 1350.8                & 645.3                 & 1440.4                & 765.0                 & 822.8                 \\ \cline{2-24} 
\multicolumn{1}{|c|}{}                                                                        & \multicolumn{1}{c|}{\multirow{3}{*}{\rotatebox{90}{General}}}  & FSST                            & 401.9                 & 376.9                 & 490.6                 & 397.6                 & 399.9                 & 347.8                 & 496.0                 & 525.2                 & 362.5                 & 559.5                 & 532.9                 & 398.0                 & 516.3                 & 431.6                 & 552.3                 & 573.2                 & 553.3                 & 362.9                 & 357.2                 & 456.2                 & 764.8                 \\
\multicolumn{1}{|c|}{}                                                                        & \multicolumn{1}{c|}{}                                                             & LZ4-d                           & \textbf{635.5}        & \textbf{608.0}        & \textbf{978.2}        & \textbf{606.8}        & \textbf{665.5}        & \textbf{665.3}        & \textbf{1108.2}       & \textbf{803.0}        & \textbf{913.0}        & {\ul \textbf{2021.5}} & \textbf{758.2}        & \textbf{740.3}        & \textbf{818.2}        & \textbf{801.2}        & {\ul \textbf{1332.9}} & \textbf{955.1}        & \textbf{2635.7}       & {\ul \textbf{3032.4}} & \textbf{2668.5}       & \textbf{3036.7}       & {\ul \textbf{2845.6}} \\
\multicolumn{1}{|c|}{}                                                                        & \multicolumn{1}{c|}{}                                                             & Zstd-d                          & 361.7                 & 367.9                 & 609.2                 & 376.7                 & 424.6                 & 314.6                 & 558.1                 & 442.9                 & 429.9                 & 919.5                 & 453.3                 & 433.1                 & 450.2                 & 413.4                 & 745.6                 & 514.7                 & 1274.3                & 1603.0                & 1481.0                & 1307.4                & 1444.5                \\ \hline
\end{tabular}
    }
\vspace{-1em}
\end{table*}
\endgroup

\subsubsection{\textbf{Baselines.}} 
We compare our methods with the following well-known lossless compression algorithms, using default compression levels and parameters for all baselines.

\begin{itemize}[left=0.07cm, topsep=0em]
    \item LZ (Lempel–Ziv) methods: The LZ family is the most widely used general-purpose compression technique, adopted by numerous database systems. Based on benchmark results~\cite{pareto_opt_compression,lzbench}, we select the following three representative algorithms as baselines:
    \begin{itemize}
        \item \textbf{LZ4}~\cite{lz4}: A lightweight compression algorithm with the fastest compression and decompression speed in the LZ family, often used in real-time scenarios.
        \item \textbf{LZMA}~\cite{lzma}: Optimized for maximum compression ratio at the cost of speed, commonly used in archival scenarios.
        \item \textbf{Zstd}~\cite{collet2018zstd}: Achieving Pareto optimality between speed and compression ratio, it is considered one of the most suitable compression methods for modern database systems~\cite{Zstdwiki}.
    \end{itemize}

    \item \textbf{FSST}~\cite{boncz2020fsst}: A state-of-the-art lightweight compression method that supports line-by-line string compression.

    \item \textbf{PBC}~\cite{zhang2023high}: A compression algorithm optimized for machine-generated data (e.g., KV, Logs, JSON), balancing general-purpose and domain-specific methods for high compression ratio and decompression speed.
    \begin{itemize}
        \item For line-by-line compression, PBC can integrate FSST for improved compression ratio, denoted as \textbf{PBC-F}.
        \item For file compression, PBC can further integrate Zstd or LZMA, denoted as \textbf{PBC-Z} and \textbf{PBC-L}, respectively.
    \end{itemize}

    \item \textbf{LogShrink}~\cite{li2024logshrink}: A log-specific compression algorithm for unstructured logs, achieving the highest compression ratio, which incorporates LZMA for further compression.

    \item \textbf{LogReducer}~\cite{yu2023logreducer}: A log-specific compression algorithm for unstructured logs, achieving Pareto optimality between compression ratio and speed, which also incorporates LZMA for further compression.

    \item \textbf{LogGrep}~\cite{wei2023loggrep}: A state-of-the-art log-specific compression algorithm for unstructured logs that supports search without full decompression. It can further integrate Zstd or LZMA, denoted as \textbf{LogGrep-Z} and \textbf{LogGrep-L}, respectively.

    \item \textbf{CLP}~\cite{rodrigues2021clp}: A state-of-the-art compression algorithm relying on pre-defined rules for unstructured and semi-structured logs, which supports search without full decompression and incorporates Zstd for further compression.

    \item \textbf{\(\mu\)Slope}~\cite{wang2024muslope}: A state-of-the-art compression algorithm for semi-structured logs, supporting search without full decompression, which incorporates Zstd for further compression.

    \item The proposed LogLite and its variations (Open-sourced at ~\cite{loglitecode}):
    \begin{itemize}
        \item \textbf{\MYLOGNAME{}-b}: A lightweight compression method proposed by us, outputting \textbf{bitstreams} and offering the best compression ratio and speed for line-by-line compression.
        \item \textbf{\MYLOGNAME{}-B}: A lightweight compression method proposed by us, outputting \textbf{bytestreams}, which can be further compressed by general-purpose compression algorithms.
        \item \textbf{\MYLOGNAME{}-BZ} and \textbf{\MYLOGNAME{}-BL}: Extensions of \MYLOGNAME{}-B with further compression using Zstd and LZMA, respectively.
    \end{itemize}
\end{itemize}

\subsection{Effectiveness and Efficiency}
Based on our study of the entire life cycle of logs, we compare \MYLOGNAME{ }with other baselines using the following most important metrics: compression ratio, compression speed, and decompression speed. 
Specifically, the compression ratio is defined as follows, where a smaller compression ratio indicates better compression:
\[
\text{Compression Ratio} = \frac{\textit{Compressed Data Size}}{\textit{Original Data Size}}
\]


In all the following experiments, unless otherwise specified, \MYLOGNAME{ }is configured with a default window size of $k = 8$ and a similarity threshold of $\theta = 0.85$. 
In Table~\ref{tab:line_compression} and \ref{tab:file_compression_json}, bold values indicate the best performance within each compressor category (log-specific or general-purpose), while bold underlined values represent the overall best across all compressors.


\subsubsection{\textbf{Line-by-Line Compression.}} 
In many applications, log producers must transmit logs to other machines or cloud servers with high real-time performance for timely analysis and monitoring. 
To ensure this, we evaluate all baseline compression algorithms supporting line-by-line compression.

According to ~\cite{boncz2020fsst}, the online compression methods in the LZ family struggle to compress single log entries. 
In most cases, their compression ratio is close to 1, or even exceeds 1, meaning no benefit can be gained from these compression methods. 
However, LZ4 and Zstd offer a special compression mode to support compression of small files. 
They first train a dictionary offline on sampled data, and then use this trained dictionary during online compression, allowing for line-by-line compression of logs. 
We evaluate LZ4 and Zstd with a pre-trained dictionary (trained by Zstd) at the default compression level, denoted \textbf{LZ4-d} and \textbf{Zstd-d}.

Notably, all baselines, including PBC, PBC-F, FSST, LZ4-d, and Zstd-d, require first collecting a sufficient amount of logs and offline training a dictionary on the sampled data to support line-by-line compression. 
In our experiment, they can sample sufficiently on each dataset. 
However, in practice, if the sampled data do not fully reflect the characteristics of the logs or if the characteristics of the logs change, the compression performance will degrade, requiring additional resources and time for resampling and retraining the dictionary. 
However, our proposed \MYLOGNAME{ }is a plug-and-play method that requires no offline training process, and can remain effective even if the characteristics of the logs change. 

As shown in Table \ref{tab:line_compression}, in terms of compression ratio, \MYLOGNAME{}-b outperforms all specific-purpose and general-purpose compression baselines. 
Compared to the second-best performer, PBC-F, \MYLOGNAME{}-b achieves an average improvement of 20.4\% in compression ratio on the TEXT dataset and 67.8\% on the JSON dataset. 
Compared to LZ4-d, the best general compressor, \MYLOGNAME{}-b achieves an average improvement of 64.7\% in compression ratio across all datasets. 

A general trend indicates that higher PSL values typically correspond to a better compression ratio for \MYLOGNAME{ }(e.g., Apa achieving a better result than Lin). In practice, the actual compression ratio is also influenced by the specific distribution of redundant characters and the encoding strategy. For example, Spa outperforms Had because the redundant characters in Spa are more contiguous, which reduces the number of RLE codes required. For datasets with a low PSL, the result is suboptimal but competitive (e.g., Ope and HDFS).

In terms of compression speed, \MYLOGNAME{}-b demonstrates strong competitiveness, achieving the best performance on the JSON dataset and the second-best performance on the TEXT dataset, only trailing FSST having a poor compression ratio. 
Compared to FSST, \MYLOGNAME{}-b achieves an average compression speed improvement of 109.5\% on the JSON dataset. 
Compared to PBC-F, \MYLOGNAME{}-b is on average 130.04\% faster across all datasets. 
For decompression speed, although LogLite is on average 41.7\% slower than PBC-F, it achieves competitive performance with FSST. 
This is because \MYLOGNAME{ }requires more computation for decompression than PBC-F's straightforward dictionary substitution.

\begingroup

\renewcommand{\arraystretch}{1.03} 
\setlength{\tabcolsep}{3.8pt}  

\begin{table}[]
\normalsize
\centering
\caption{File compression performance for semi-structured (JSON) log. (Speed is measured in MB/s)}
\vspace{-0.5em}
\label{tab:file_compression_json}
\resizebox{0.90\columnwidth}{!}{%
\begin{tabular}{|ccc|ccccc|}
\hline
\multicolumn{3}{|c|}{Datasets}                                                                                                                                                                                       & Mon                    & Coc                    & Ela                    & Spark                  & Pos                    \\ \hline
\multicolumn{1}{|c|}{\multirow{9}{*}{\rotatebox{90}{Compression Ratio}}}   & \multicolumn{1}{c|}{\multirow{6}{*}{\rotatebox{90}{Specific}}} & CLP                              & 0.0091                 & 0.0497                 & 0.0073                 & 0.0209                 & 0.0328                 \\
\multicolumn{1}{|c|}{}                                                                        & \multicolumn{1}{c|}{}                                                             & \(\mu\)Slope                     & 0.0059                 & 0.0412                 & 0.0067                 & 0.0199                 & 0.0510                 \\
\multicolumn{1}{|c|}{}                                                                        & \multicolumn{1}{c|}{}                                                             & PBC-Z                            & 0.0253                 & 0.0445                 & 0.0074                 & 0.0265                 & 0.0238                 \\
\multicolumn{1}{|c|}{}                                                                        & \multicolumn{1}{c|}{}                                                             & PBC-L                            & 0.0152                 & {\ul \textbf{0.0258}}  & 0.0042                 & 0.0185                 & {\ul \textbf{0.0161}}  \\
\multicolumn{1}{|c|}{}                                                                        & \multicolumn{1}{c|}{}                                                             & \MYLOGNAME{}-BZ & 0.0045                 & 0.0398                 & 0.0037                 & 0.0240                 & 0.0308                 \\
\multicolumn{1}{|c|}{}                                                                        & \multicolumn{1}{c|}{}                                                             & \MYLOGNAME{}-BL & {\ul \textbf{0.0026}}  & 0.0284                 & {\ul \textbf{0.0018}}  & {\ul \textbf{0.0179}}  & 0.0225                 \\ \cline{2-8} 
\multicolumn{1}{|c|}{}                                                                        & \multicolumn{1}{c|}{\multirow{3}{*}{\rotatebox{90}{General}}}  & LZ4                              & 0.0388                 & 0.0907                 & 0.0325                 & 0.1601                 & 0.0681                 \\
\multicolumn{1}{|c|}{}                                                                        & \multicolumn{1}{c|}{}                                                             & Zstd                             & 0.0140                 & 0.0485                 & \textbf{0.0107}        & 0.0359                 & 0.0434                 \\
\multicolumn{1}{|c|}{}                                                                        & \multicolumn{1}{c|}{}                                                             & LZMA                             & \textbf{0.0138}        & \textbf{0.0343}        & 0.0114                 & \textbf{0.0241}        & \textbf{0.0343}        \\ \hline
\multicolumn{1}{|c|}{\multirow{9}{*}{\rotatebox{90}{Compression Speed}}}   & \multicolumn{1}{c|}{\multirow{6}{*}{\rotatebox{90}{Specific}}} & CLP                              & 93.15                  & 45.57                  & 113.94                 & 104.47                 & 57.54                  \\
\multicolumn{1}{|c|}{}                                                                        & \multicolumn{1}{c|}{}                                                             & \(\mu\)Slope                     & 89.37                  & 54.68                  & 103.84                 & 91.28                  & 68.79                  \\
\multicolumn{1}{|c|}{}                                                                        & \multicolumn{1}{c|}{}                                                             & PBC-Z                            & \textbf{349.48}        & \textbf{224.35}        & 405.34                 & 170.46                 & 155.30                 \\
\multicolumn{1}{|c|}{}                                                                        & \multicolumn{1}{c|}{}                                                             & PBC-L                            & 24.17                  & 10.17                  & 44.78                  & 12.61                  & 19.19                  \\
\multicolumn{1}{|c|}{}                                                                        & \multicolumn{1}{c|}{}                                                             & \MYLOGNAME{}-BZ & 328.90                 & 167.84                 & \textbf{444.66}        & \textbf{288.73}        & \textbf{254.90}        \\
\multicolumn{1}{|c|}{}                                                                        & \multicolumn{1}{c|}{}                                                             & \MYLOGNAME{}-BL & 106.01                 & 12.33                  & 123.70                 & 23.17                  & 21.53                  \\ \cline{2-8} 
\multicolumn{1}{|c|}{}                                                                        & \multicolumn{1}{c|}{\multirow{3}{*}{\rotatebox{90}{General}}}  & LZ4                              & {\ul \textbf{4386.99}} & {\ul \textbf{1754.00}} & 2997.00                & 997.00                 & {\ul \textbf{2890.00}} \\
\multicolumn{1}{|c|}{}                                                                        & \multicolumn{1}{c|}{}                                                             & Zstd                             & 2579.15                & 859.00                 & {\ul \textbf{3095.00}} & {\ul \textbf{1131.00}} & 1272.00                \\
\multicolumn{1}{|c|}{}                                                                        & \multicolumn{1}{c|}{}                                                             & LZMA                             & 22.55                  & 12.60                  & 25.80                  & 9.26                   & 15.30                  \\ \hline
\multicolumn{1}{|c|}{\multirow{9}{*}{\rotatebox{90}{Decompression Speed}}} & \multicolumn{1}{c|}{\multirow{6}{*}{\rotatebox{90}{Specific}}} & CLP                              & 160.08                 & 78.22                  & 666.57                 & 368.10                 & 384.76                 \\
\multicolumn{1}{|c|}{}                                                                        & \multicolumn{1}{c|}{}                                                             & \(\mu\)Slope                     & 185.45                 & 216.78                 & 491.87                 & 290.10                 & 127.96                 \\
\multicolumn{1}{|c|}{}                                                                        & \multicolumn{1}{c|}{}                                                             & PBC-Z                            & \textbf{1784.62}       & \textbf{1892.38}       & \textbf{2865.79}       & \textbf{2268.43}       & \textbf{2014.68}       \\
\multicolumn{1}{|c|}{}                                                                        & \multicolumn{1}{c|}{}                                                             & PBC-L                            & 961.79                 & 454.97                 & 1586.72                & 828.58                 & 839.34                 \\
\multicolumn{1}{|c|}{}                                                                        & \multicolumn{1}{c|}{}                                                             & \MYLOGNAME{}-BZ & 1263.38                & 571.33                 & 1215.86                & 718.83                 & 822.42                 \\
\multicolumn{1}{|c|}{}                                                                        & \multicolumn{1}{c|}{}                                                             & \MYLOGNAME{}-BL & 1157.86                & 297.41                 & 1136.19                & 434.61                 & 455.88                 \\ \cline{2-8} 
\multicolumn{1}{|c|}{}                                                                        & \multicolumn{1}{c|}{\multirow{3}{*}{\rotatebox{90}{General}}}  & LZ4                              & {\ul \textbf{5680.40}} & {\ul \textbf{6722.00}} & {\ul \textbf{7046.00}} & {\ul \textbf{5046.00}} & {\ul \textbf{7968.00}} \\
\multicolumn{1}{|c|}{}                                                                        & \multicolumn{1}{c|}{}                                                             & Zstd                             & 4581.71                & 2545.00                & 6605.00                & 3427.00                & 3031.00                \\
\multicolumn{1}{|c|}{}                                                                        & \multicolumn{1}{c|}{}                                                             & LZMA                             & 987.50                 & 480.00                 & 831.00                 & 775.00                 & 615.00                 \\ \hline
\end{tabular}
}
\vspace{-1.5em}
\end{table}

\endgroup

\begingroup

\renewcommand{\arraystretch}{1.00} 
\setlength{\tabcolsep}{3.1pt}  

\begin{table*}[t]
    \centering
\caption{File compression performance for unstructured (Text) logs. (Speed is measured in MB/s)}
\label{tab:file_compression_text}
    \resizebox{\textwidth}{!}{%
\begin{tabular}{|ccc|cccccccccccccccc|}
\hline
\multicolumn{3}{|c|}{Datasets}                                                                                                                                                                                        & Apa                   & Lin                   & Zoo                   & Hea                   & HPC                   & And                   & Had                   & BGL                   & Mac                   & Ope                   & Pro                   & Spa                   & SSH                   & Thu                   & Win                   & HDFS                  \\ \hline
\multicolumn{1}{|c|}{\multirow{12}{*}{\rotatebox{90}{Compression Ratio}}}   & \multicolumn{1}{c|}{\multirow{9}{*}{\rotatebox{90}{Specific}}} & LogReducer                       & 0.023                 & 0.036                 & 0.009                 & 0.068                 & 0.035                 & 0.043                 & 0.018                 & 0.034                 & 0.026                 & {\ul \textbf{0.042}}  & {\ul \textbf{0.0363}} & 0.020                 & 0.012                 & {\ul \textbf{0.017}}  & 0.0029                & 0.043                 \\
\multicolumn{1}{|c|}{}                                                                         & \multicolumn{1}{c|}{}                                                             & LogShrink                        & {\ul \textbf{0.019}}  & {\ul \textbf{0.034}}  & {\ul \textbf{0.008}}  & 0.072                 & {\ul \textbf{0.025}}  & 0.043                 & 0.018                 & {\ul \textbf{0.027}}  & {\ul \textbf{0.025}}  & 0.043                 & 0.0364                & 0.019                 & {\ul \textbf{0.009}}  & 0.018                 & 0.0021                & {\ul \textbf{0.035}}  \\
\multicolumn{1}{|c|}{}                                                                         & \multicolumn{1}{c|}{}                                                             & LogGrep-Z                        & 0.028                 & 0.064                 & 0.025                 & 0.062                 & 0.056                 & 0.058                 & 0.019                 & 0.035                 & 0.043                 & 0.082                 & 0.0565                & 0.034                 & 0.038                 & 0.029                 & 0.0030                & 0.064                 \\
\multicolumn{1}{|c|}{}                                                                         & \multicolumn{1}{c|}{}                                                             & LogGrep-L                        & 0.023                 & 0.048                 & 0.019                 & {\ul \textbf{0.041}}  & 0.032                 & {\ul \textbf{0.036}}  & {\ul \textbf{0.016}}  & 0.027                 & 0.036                 & 0.052                 & 0.0500                & {\ul \textbf{0.018}}  & 0.021                 & 0.024                 & 0.0024                & 0.052                 \\
\multicolumn{1}{|c|}{}                                                                         & \multicolumn{1}{c|}{}                                                             & CLP                              & 0.068                 & 0.102                 & 0.030                 & 0.081                 & 0.066                 & 0.049                 & 0.037                 & 0.090                 & 0.076                 & 0.077                 & 0.1065                & 0.047                 & 0.056                 & 0.053                 & 0.0022                & 0.061                 \\
\multicolumn{1}{|c|}{}                                                                         & \multicolumn{1}{c|}{}                                                             & PBC-Z                            & 0.038                 & 0.062                 & 0.023                 & 0.080                 & 0.086                 & 0.059                 & 0.028                 & 0.079                 & 0.045                 & 0.063                 & 0.0534                & 0.049                 & 0.048                 & 0.051                 & 0.0045                & 0.076                 \\
\multicolumn{1}{|c|}{}                                                                         & \multicolumn{1}{c|}{}                                                             & PBC-L                            & 0.026                 & 0.039                 & 0.013                 & 0.049                 & 0.049                 & 0.038                 & 0.019                 & 0.041                 & 0.032                 & 0.052                 & 0.0417                & 0.032                 & 0.027                 & 0.035                 & 0.0026                & 0.053                 \\
\multicolumn{1}{|c|}{}                                                                         & \multicolumn{1}{c|}{}                                                             & \MYLOGNAME{}-BZ & 0.044                 & 0.068                 & 0.024                 & 0.082                 & 0.065                 & 0.078                 & 0.033                 & 0.056                 & 0.050                 & 0.071                 & 0.0716                & 0.035                 & 0.042                 & 0.053                 & 0.0038                & 0.097                 \\
\multicolumn{1}{|c|}{}                                                                         & \multicolumn{1}{c|}{}                                                             & \MYLOGNAME{}-BL & 0.036                 & 0.055                 & 0.018                 & 0.064                 & 0.052                 & 0.051                 & 0.025                 & 0.041                 & 0.040                 & 0.061                 & 0.0601                & 0.024                 & 0.031                 & 0.031                 & {\ul \textbf{0.0012}} & 0.072                 \\ \cline{2-19} 
\multicolumn{1}{|c|}{}                                                                         & \multicolumn{1}{c|}{\multirow{3}{*}{\rotatebox{90}{General}}}  & LZ4                              & 0.088                 & 0.158                 & 0.083                 & 0.167                 & 0.177                 & 0.197                 & 0.082                 & 0.167                 & 0.137                 & 0.144                 & 0.1149                & 0.123                 & 0.115                 & 0.103                 & 0.0891                & 0.176                 \\
\multicolumn{1}{|c|}{}                                                                         & \multicolumn{1}{c|}{}                                                             & Zstd                             & 0.052                 & 0.081                 & 0.043                 & 0.100                 & 0.102                 & 0.075                 & 0.038                 & 0.094                 & 0.058                 & 0.080                 & 0.0730                & 0.065                 & 0.070                 & 0.052                 & 0.0066                & 0.096                 \\
\multicolumn{1}{|c|}{}                                                                         & \multicolumn{1}{c|}{}                                                             & LZMA                             & \textbf{0.039}        & \textbf{0.059}        & \textbf{0.036}        & \textbf{0.075}        & \textbf{0.067}        & \textbf{0.053}        & \textbf{0.028}        & \textbf{0.058}        & \textbf{0.045}        & \textbf{0.067}        & \textbf{0.0527}       & \textbf{0.050}        & \textbf{0.054}        & \textbf{0.037}        & \textbf{0.0042}       & \textbf{0.074}        \\ \hline
\multicolumn{1}{|c|}{\multirow{12}{*}{\rotatebox{90}{Compression Speed}}}   & \multicolumn{1}{c|}{\multirow{9}{*}{\rotatebox{90}{Specific}}} & LogReducer                       & 3.1                   & 1.8                   & 5.6                   & 3.3                   & 4.0                   & 7.4                   & 5.1                   & 7.6                   & 4.5                   & 7.8                   & 1.9                   & 9.7                   & 5.6                   & 8.9                   & 14.2                  & 7.4                   \\
\multicolumn{1}{|c|}{}                                                                         & \multicolumn{1}{c|}{}                                                             & LogShrink                        & 1.4                   & 0.6                   & 1.9                   & 3.9                   & 1.8                   & 1.0                   & 2.0                   & 1.1                   & 1.1                   & 2.0                   & 0.5                   & 1.1                   & 1.7                   & 0.2                   & 1.2                   & 1.3                   \\
\multicolumn{1}{|c|}{}                                                                         & \multicolumn{1}{c|}{}                                                             & LogGrep-Z                        & 25.2                  & 13.4                  & 35.9                  & 26.1                  & 30.5                  & 17.2                  & 50.2                  & 24.3                  & 31.7                  & 24.1                  & 18.8                  & 25.8                  & 27.5                  & 26.6                  & 62.8                  & 30.5                  \\
\multicolumn{1}{|c|}{}                                                                         & \multicolumn{1}{c|}{}                                                             & LogGrep-L                        & 9.7                   & 5.6                   & 14.3                  & 2.8                   & 4.3                   & 6.9                   & 12.9                  & 8.2                   & 8.4                   & 10.4                  & 0.8                   & 4.2                   & 5.3                   & 9.2                   & 25.2                  & 6.1                   \\
\multicolumn{1}{|c|}{}                                                                         & \multicolumn{1}{c|}{}                                                             & CLP                              & 2.7                   & 1.1                   & 3.9                   & 8.7                   & 11.4                  & 40.0                  & 14.0                  & 29.5                  & 7.1                   & 7.1                   & 1.3                   & 73.8                  & 21.6                  & 52.1                  & 120.6                 & 46.4                  \\
\multicolumn{1}{|c|}{}                                                                         & \multicolumn{1}{c|}{}                                                             & PBC-Z                            & 51.7                  & 63.6                  & 122.2                 & 55.9                  & 59.6                  & 49.4                  & 95.9                  & 31.1                  & 87.1                  & \textbf{156.3}        & 31.8                  & 49.7                  & 54.1                  & 105.1                 & 104.3                 & 83.9                  \\
\multicolumn{1}{|c|}{}                                                                         & \multicolumn{1}{c|}{}                                                             & PBC-L                            & 19.6                  & 17.2                  & 32.6                  & 9.5                   & 9.8                   & 6.8                   & 21.3                  & 4.8                   & 15.0                  & 16.4                  & 15.8                  & 7.0                   & 11.3                  & 8.6                   & 32.8                  & 3.7                   \\
\multicolumn{1}{|c|}{}                                                                         & \multicolumn{1}{c|}{}                                                             & \MYLOGNAME{}-BZ & \textbf{94.2}         & \textbf{72.0}         & \textbf{142.6}        & \textbf{73.8}         & \textbf{77.4}         & \textbf{76.8}         & \textbf{113.2}        & \textbf{120.7}        & \textbf{97.4}         & 123.9                 & \textbf{84.7}         & \textbf{100.4}        & \textbf{103.4}        & \textbf{124.9}        & \textbf{158.1}        & \textbf{81.4}         \\
\multicolumn{1}{|c|}{}                                                                         & \multicolumn{1}{c|}{}                                                             & \MYLOGNAME{}-BL & 29.9                  & 22.3                  & 42.9                  & 16.3                  & 17.3                  & 9.2                   & 33.1                  & 7.8                   & 25.0                  & 17.5                  & 24.4                  & 12.6                  & 18.7                  & 12.3                  & 69.8                  & 3.5                   \\ \cline{2-19} 
\multicolumn{1}{|c|}{}                                                                         & \multicolumn{1}{c|}{\multirow{3}{*}{\rotatebox{90}{General}}}  & LZ4                              & {\ul \textbf{1883.0}} & {\ul \textbf{1178.0}} & {\ul \textbf{2331.0}} & {\ul \textbf{985.0}}  & {\ul \textbf{1185.0}} & {\ul \textbf{874.0}}  & {\ul \textbf{1951.0}} & {\ul \textbf{1214.0}} & {\ul \textbf{1225.0}} & {\ul \textbf{1380.0}} & {\ul \textbf{1387.0}} & {\ul \textbf{1646.0}} & {\ul \textbf{1889.0}} & {\ul \textbf{1592.0}} & 2293.7                & {\ul \textbf{1103.0}} \\
\multicolumn{1}{|c|}{}                                                                         & \multicolumn{1}{c|}{}                                                             & Zstd                             & 1128.0                & 726.0                 & 1281.0                & 530.0                 & 552.0                 & 658.0                 & 1330.0                & 525.0                 & 909.0                 & 599.0                 & 836.0                 & 790.0                 & 780.0                 & 914.1                 & {\ul \textbf{3333.6}} & 535.0                 \\
\multicolumn{1}{|c|}{}                                                                         & \multicolumn{1}{c|}{}                                                             & LZMA                             & 13.7                  & 9.5                   & 13.6                  & 6.3                   & 5.8                   & 8.2                   & 13.1                  & 6.1                   & 11.2                  & 8.7                   & 11.5                  & 7.6                   & 8.4                   & 10.9                  & 21.0                  & 5.7                   \\ \hline
\multicolumn{1}{|c|}{\multirow{10}{*}{\rotatebox{90}{Decompression Speed}}} & \multicolumn{1}{c|}{\multirow{7}{*}{\rotatebox{90}{Specific}}} & LogReducer                       & 7.2                   & 2.6                   & 5.2                   & 5.7                   & 6.2                   & 2.6                   & 8.6                   & 8.2                   & 4.6                   & 10.9                  & 1.5                   & 7.0                   & 6.5                   & 1.7                   & 12.1                  & 10.0                  \\
\multicolumn{1}{|c|}{}                                                                         & \multicolumn{1}{c|}{}                                                             & LogShrink                        & 4.4                   & 1.7                   & 4.9                   & 9.8                   & 5.3                   & 2.1                   & 5.6                   & 3.5                   & 1.8                   & 5.8                   & 1.2                   & 5.1                   & 4.3                   & 0.6                   & 6.2                   & 4.9                   \\
\multicolumn{1}{|c|}{}                                                                         & \multicolumn{1}{c|}{}                                                             & CLP                              & 86.8                  & 48.3                  & 142.0                 & 112.2                 & 138.5                 & 189.0                 & 181.1                 & 162.5                 & 65.5                  & 127.5                 & 4.5                   & 194.1                 & 95.4                  & 163.6                 & 386.4                 & 268.5                 \\
\multicolumn{1}{|c|}{}                                                                         & \multicolumn{1}{c|}{}                                                             & PBC-Z                            & \textbf{1775.7}       & \textbf{1492.0}       & \textbf{1956.5}       & \textbf{1207.6}       & \textbf{1167.7}       & \textbf{1234.6}       & \textbf{2137.6}       & \textbf{918.8}        & \textbf{1128.8}       & \textbf{1506.0}       & \textbf{1450.8}       & \textbf{1287.9}       & \textbf{1642.2}       & \textbf{704.4}        & \textbf{1001.8}       & \textbf{1241.8}       \\
\multicolumn{1}{|c|}{}                                                                         & \multicolumn{1}{c|}{}                                                             & PBC-L                            & 528.7                 & 372.9                 & 839.2                 & 293.0                 & 275.2                 & 399.8                 & 721.0                 & 287.0                 & 421.1                 & 296.6                 & 368.8                 & 462.4                 & 502.1                 & 324.6                 & 953.7                 & 324.0                 \\
\multicolumn{1}{|c|}{}                                                                         & \multicolumn{1}{c|}{}                                                             & \MYLOGNAME{}-BZ & 471.4                 & 356.3                 & 609.0                 & 322.2                 & 383.5                 & 298.3                 & 509.2                 & 524.8                 & 383.6                 & 440.6                 & 390.2                 & 489.5                 & 475.4                 & 525.6                 & 719.3                 & 339.3                 \\
\multicolumn{1}{|c|}{}                                                                         & \multicolumn{1}{c|}{}                                                             & \MYLOGNAME{}-BL & 269.5                 & 187.6                 & 407.3                 & 166.4                 & 201.7                 & 175.4                 & 325.0                 & 260.5                 & 226.6                 & 194.9                 & 189.9                 & 342.0                 & 286.4                 & 318.6                 & 696.6                 & 169.2                 \\ \cline{2-19} 
\multicolumn{1}{|c|}{}                                                                         & \multicolumn{1}{c|}{\multirow{3}{*}{\rotatebox{90}{General}}}  & LZ4                              & {\ul \textbf{5905.0}} & {\ul \textbf{5122.0}} & {\ul \textbf{7428.0}} & {\ul \textbf{3916.0}} & {\ul \textbf{4999.0}} & {\ul \textbf{3801.0}} & {\ul \textbf{6992.0}} & {\ul \textbf{4633.0}} & {\ul \textbf{4828.0}} & {\ul \textbf{6592.0}} & {\ul \textbf{5087.0}} & {\ul \textbf{4287.0}} & {\ul \textbf{6026.0}} & {\ul \textbf{3646.0}} & {\ul \textbf{8120.8}} & {\ul \textbf{4319.0}} \\
\multicolumn{1}{|c|}{}                                                                         & \multicolumn{1}{c|}{}                                                             & Zstd                             & 2872.0                & 2194.0                & 3302.0                & 1743.0                & 1803.0                & 2013.0                & 3718.0                & 1583.0                & 2605.0                & 2910.0                & 2464.0                & 1885.0                & 2322.0                & 1927.8                & 6442.1                & 1727.0                \\
\multicolumn{1}{|c|}{}                                                                         & \multicolumn{1}{c|}{}                                                             & LZMA                             & 465.0                 & 318.0                 & 557.0                 & 270.0                 & 279.0                 & 377.0                 & 704.0                 & 247.0                 & 436.0                 & 297.0                 & 364.0                 & 280.0                 & 387.0                 & 417.6                 & 1074.3                & 297.0                 \\ \hline
\end{tabular}
    }
\end{table*}
\endgroup

\input{tables/pareto}
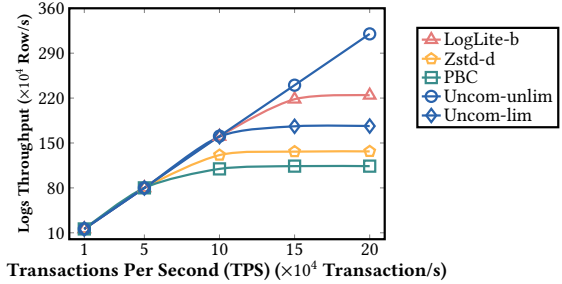
\begin{figure}
  \centering
  \begin{tikzpicture}[scale=0.42]
    \begin{axis}[
        legend style={
            at={(1.1,0.7)},      
            anchor=west,          
            font=\Huge,           
            draw=black,           
            legend cell align=left,  
        },
        width=0.65\textwidth,
        height=0.5\textwidth,
        ymin=0,ymax=360,
        ytick = {10,80,150,220,290,360},
        mark size=4.0pt,
        ticklabel style={align=center,font=\Huge},
        xtick = {1,5,10,15,20},
        xmin=0,xmax=21,
        xlabel={\Huge \bf Transactions Per Second (TPS) ($\times 10^4$ Transaction/s)},
        ylabel={\huge \bf Logs Throughput ($\times 10^4$ Row/s)},
        ylabel style={yshift=0.2cm},
        every axis plot/.append style={line width = 2pt},
        every axis/.append style={line width = 2pt},
    ]

    \addplot[smooth,mark=triangle,color=c5,mark size=6pt] plot coordinates { 
      (1,16.0039)
      (5,80.0090)
      (10,160.1757)
      (15,218.2159)
      (20,224.3570)
    };

    \addplot[smooth,mark=pentagon,color=c3,mark size=5pt] plot coordinates { 
      (1,16.0059)
      (5,79.9556)
      (10,131.1018)
      (15,136.5267)
      (20,136.8642)
    };

    \addplot[smooth,mark=square,color=c8,mark size=5pt] plot coordinates { 
      (1,16.0112)
      (5,80.0598)
      (10,109.6306)
      (15,113.9332)
      (20,113.8980)
    };
    
    \addplot[smooth,mark=o,color=c1,mark size=5pt] plot coordinates { 
      (1,16.0386)
      (5,80.0085)
      (10,160.6901)
      (15,240.1144)
      (20,320.0796)
    };

    \addplot[smooth,mark=diamond,color=c1,mark size=6pt] plot coordinates { 
      (1,15.9787)
      (5,79.8865)
      (10,159.8799)
      (15,176.0016)
      (20,176.3701)
    };

    \legend{\MYLOGNAME{}-b,Zstd-d,PBC,Uncom-unlim,Uncom-lim}
    \end{axis}
  \end{tikzpicture}

  \vspace{-3mm}
  \caption{Throughput of logs processing.}
  \label{fig:casestudy}
  \vspace{-1.5em}
\end{figure}

\subsubsection{\textbf{File Compression.}} 
Before logs are archived to disk, they are typically collected and compressed as a large file, significantly reducing storage costs. 
This approach is particularly well-suited for the LZ family, as these algorithms excel at finding redundancy over large blocks of data. 
This is also a common application scenario for many log-purpose compression baselines. 
Although \MYLOGNAME{ }first compresses log entries on line-by-line, thanks to our design, \MYLOGNAME{ }exhibits a degree of orthogonality with the LZ family. 
As a result, algorithms like Zstd and LZMA can still further reduce redundancies effectively. 
Specifically, Zstd and LZMA can perform additional compression on the data files produced by \MYLOGNAME{}-B’s line-by-line compression (i.e., \MYLOGNAME{}-BZ and \MYLOGNAME{}-BL). 
Most log-purpose compression baselines support only one type of log format: either unstructured (TEXT) or semi-structured (JSON), but our method is applicable to both common log formats.

In the JSON datasets, as shown in Table \ref{tab:file_compression_json}, \MYLOGNAME{}-BL achieves the best compression ratio, outperforming the second-best method, PBC-L, by 8.1\% on average, and the best general-purpose compressor, LZMA, by 37.7\% on average. 
Additionally, \MYLOGNAME{}-BZ surpasses PBC-Z and \(\mu\)Slope by an average of 19.2\% and 17.3\%, respectively. 
In terms of compression speed, \MYLOGNAME{}-BZ achieves the best results among specific-purpose methods, outperforming PBC-Z by an average of 14.2\%. 
Similarly, \MYLOGNAME{}-BL outpaces PBC-L and LZMA by an average of 1.6× and 2.4×, respectively. 
For decompression speed, although \MYLOGNAME{}-L is on average 25.5\% slower than PBC-L, it demonstrates competitive overall performance.




In the TEXT datasets, as shown in Table \ref{tab:file_compression_text}, some specific compressors sacrifice speed for better compression by using complex computations and more resources to exploit unstructured log patterns. 
Despite this, \MYLOGNAME{}-BL still achieves a competitive compression ratio, outperforming LZMA by an average of 17.1\%. 
Among specific-purpose methods, \MYLOGNAME{}-BZ achieves the fastest compression speed, outperforming LogGrep-Z and PBC-Z on average 2.5× and 36.9\%, respectively.
\MYLOGNAME{}-BL demonstrates a compression speed that is, on average, 15×, 2.7×, 1.7× and 56.3\% faster than the compression ratio leaders LogShrink\footnote{Although LogShrink ~\cite{li2024logshrink} suggests a C++ implementation could be 10x faster than its Python version, LogLite still holds a speed advantage, especially in decompression.}, LogReducer, LogGrep-L and PBC-L, respectively, and 1.24× faster than LZMA. 
For decompression speed\footnote{We don't test LogGrep's decompression as its decompression code is not open-source.}, although \MYLOGNAME{}-L is on average 40.06\% slower than PBC-L, it still outperforms LogShrink and LogReducer by two orders of magnitude and achieves a speed comparable to LZMA. 

\subsubsection{\textbf{Pareto-Optimal Compression.}} 
\label{sec:pareto_compression}
To better illustrate the position of \MYLOGNAME{ }in compressing TEXT and JSON logs, we plot 8 figures based on the aforementioned experimental results over all our TEXT and JSON log datasets. 
These figures respectively depict the positions of compression ratio versus compression speed and compression ratio versus decompression speed for both line-by-line and file compression on TEXT and JSON logs.
The top-left corner of each graph indicates better performance, and the red line represents the Pareto frontier. 
Methods positioned on this red line are Pareto-optimal. 
To distinguish \MYLOGNAME{ }from other baselines, we depict it as an upward triangle in the plots. 
As shown in Figure \ref{fig:pareto}, \MYLOGNAME{ }achieves Pareto-Optimal in seven out of the eight figures. 
Notably, for line-by-line compression on JSON logs, \MYLOGNAME{ }achieves the best results in both compression ratio and compression speed.

\begin{figure*}
\centering
\subfigure[Compression ratio]{
    \begin{tikzpicture}[scale=0.37]
\begin{axis}[
    ybar=0pt, 
    bar width=0.35cm, 
    scaled y ticks=false,
    width=0.65\textwidth, 
    height=0.5\textwidth, 
    xtick=data,  
    xticklabels={\Huge $Apa$, \Huge $And$, \Huge $Win$, \Huge $Ela$, \Huge $Pos$},
    legend style={at={(0.5,0.98)},
        anchor=north, legend columns=-1,
        draw=none},
    legend image code/.code={
        \draw [#1] (0cm,-0.263cm) rectangle (1.2cm,0.085cm); },
    ymin=0, ymax=0.09, 
    ytick={0,0.02,0.04,0.06,0.08}, 
    yticklabels={0,0.02,0.04,0.06,0.08},
    xmin=0.5,xmax=5.5,
    tick align=inside,
    ticklabel style={font=\Huge},
    every axis plot/.append style={line width=2pt},
    every axis/.append style={line width=2pt},
    ylabel={\textbf{\huge Compression Ratio}},
    ylabel style={yshift=0.2cm}
]

                \addplot [pattern=myDenseLines, pattern color=c1, fill opacity=1] coordinates {
(1, 0.0349643)
(2, 0.0504499)
(3, 0.00120291)
(4, 0.00181444)
(5, 0.023053)
                };
                \addplot [pattern=myDenseLinesRev,pattern color=c8, fill opacity=1] coordinates {
(1, 0.0427575)
(2, 0.0554389)
(3, 0.00220553)
(4, 0.00305032)
(5, 0.0266113)
                };
                \addplot [pattern=myCrosshatch ,pattern color=American Yellow, fill opacity=1] coordinates {
(1, 0.038644)
(2, 0.0747763)
(3, 0.0015685)
(4, 0.00185272)
(5, 0.0240118)
                };
              \addplot [pattern=myHorizontalLines , pattern color=c5, fill opacity=1] coordinates {
(1, 0.049563)
(2, 0.0717779)
(3, 0.00380342)
(4, 0.00249993)
(5, 0.0289509)
                }; 
        \end{axis}
        \label{fig:ablation1}
      \end{tikzpicture}
}  
\subfigure[Compression speed]{
    \begin{tikzpicture}[scale=0.37]
\begin{axis}[
    ybar=0pt, 
    bar width=0.35cm, 
    scaled y ticks=false,
    width=0.65\textwidth, 
    height=0.5\textwidth, 
    xtick=data,  
    xticklabels={\Huge $Apa$, \Huge $And$, \Huge $Win$, \Huge $Ela$, \Huge $Pos$},
    legend style={at={(0.5,0.98)},
        anchor=north, legend columns=-1,
        draw=none},
    legend image code/.code={
        \draw [#1] (0cm,-0.263cm) rectangle (1.2cm,0.085cm); },
    ymin=0, ymax=125, 
    ytick={0,25,50,75,100,125}, 
    yticklabels={0,25,50,75,100,125},
    xmin=0.5,xmax=5.5,
    tick align=inside,
    ticklabel style={font=\Huge},
    every axis plot/.append style={line width=2pt},
    every axis/.append style={line width=2pt},
    ylabel={\textbf{\huge Compression Speed (MB/s)}},
    ylabel style={yshift=0.2cm}
]

                \addplot [pattern=myDenseLines, pattern color=c1, fill opacity=1] coordinates {
(1, 25.4462)
(2, 8.7499)
(3, 68.4942)
(4, 110.016)
(5, 18.4592)
                };
                \addplot [pattern=myDenseLinesRev ,pattern color=c8, fill opacity=1] coordinates {
(1, 22.1265)
(2, 7.97601)
(3, 55.4344)
(4, 68.1245)
(5, 16.5312)
                };
                \addplot [pattern=myCrosshatch ,pattern color=American Yellow, fill opacity=1] coordinates {
(1, 24.973)
(2, 7.26831)
(3, 66.867)
(4, 112.347)
(5, 20.0005)
                };
              \addplot [pattern=myHorizontalLines , pattern color=c5, fill opacity=1] coordinates {
(1, 32.3269)
(2, 6.32818)
(3, 51.1483)
(4, 51.1483)
(5, 19.8131)
                }; 
        \end{axis}
        \label{fig:ablation2}
      \end{tikzpicture}
}
\subfigure[Decompression speed]{
    \begin{tikzpicture}[scale=0.37]
\begin{axis}[
    ybar=0pt, 
    bar width=0.35cm, 
    scaled y ticks=false,
    width=0.65\textwidth, 
    height=0.5\textwidth, 
    xtick=data,  
    xticklabels={\Huge $Apa$, \Huge $And$, \Huge $Win$, \Huge $Ela$, \Huge $Pos$},
    legend style={at={(0.5,0.98)},
        anchor=north, legend columns=-1,
        draw=none},
    legend image code/.code={
        \draw [#1] (0cm,-0.263cm) rectangle (1.2cm,0.085cm); },
    ymin=0, ymax=1500, 
    ytick={0,500,1000,1500}, 
    yticklabels={0,500,1000,1500},
    xmin=0.5,xmax=5.5,
    tick align=inside,
    ticklabel style={font=\Huge},
    every axis plot/.append style={line width=2pt},
    every axis/.append style={line width=2pt},
    ylabel={\textbf{\huge Decompression Speed (MB/s)}},
    ylabel style={yshift=0.2cm}
]

                \addplot [pattern=myDenseLines, pattern color=c1, fill opacity=1] coordinates {
(1, 268.597)
(2, 180.869)
(3, 638.067)
(4, 1189.79)
(5, 418.754)
                };
                \addplot [pattern=myDenseLinesRev ,pattern color=c8, fill opacity=1] coordinates {
(1, 221.5)
(2, 170.289)
(3, 601.837)
(4, 1048.16)
(5, 395.406)
                };
                \addplot [pattern=myCrosshatch ,pattern color=American Yellow, fill opacity=1] coordinates {
(1, 251.814)
(2, 148.877)
(3, 633.536)
(4, 1185.45)
(5, 416.852)
                };
              \addplot [pattern=myHorizontalLines , pattern color=c5, fill opacity=1] coordinates {
(1, 241.383)
(2, 163.755)
(3, 638.001)
(4, 1155.37)
(5, 430.501)
                }; 
        \end{axis}
        \label{fig:ablation3}
      \end{tikzpicture}%
}
\hspace*{0.5cm}
\raisebox{0.5cm}{
\begin{tikzpicture}[scale=0.5]
  \begin{axis}[
    axis line style=none, 
    scale only axis,
    height=0pt, 
    width=0pt,  
    xmin=0, xmax=1, ymin=0, ymax=1, 
    legend style={
        at={(0,1)}, 
        anchor=north, 
        legend columns=1, 
        minimum width=3.15cm, 
        minimum height=1cm, 
        column sep=0.1em, 
        draw=black, 
        legend cell align={left}, 
        inner xsep=5pt, 
        inner ysep=2pt 
    },
    legend image code/.code={
        \draw [#1] (-0.5cm,-0.2cm) rectangle (0.5cm,0.2cm); 
    },
  ]
    \addlegendimage{pattern=myDenseLines, pattern color=c1}
    \addlegendentry{\huge \MYLOGNAME{}-BL } 
    
    \addlegendimage{pattern=myDenseLinesRev, pattern color=c8}
    \addlegendentry{\huge w/o Reverse } 
  
    \addlegendimage{pattern=myCrosshatch, pattern color=c3}
    \addlegendentry{\huge w/o XOR-P } 

    \addlegendimage{pattern=myHorizontalLines, pattern color=c5}
    \addlegendentry{\huge w/o RLE-B } 
  \end{axis}
\end{tikzpicture}
}
  \vspace{-3mm}
    \caption{Ablation study on 5 representative datasets.}
    \vspace{-1em}
\label{fig:ablation}
      
\end{figure*}
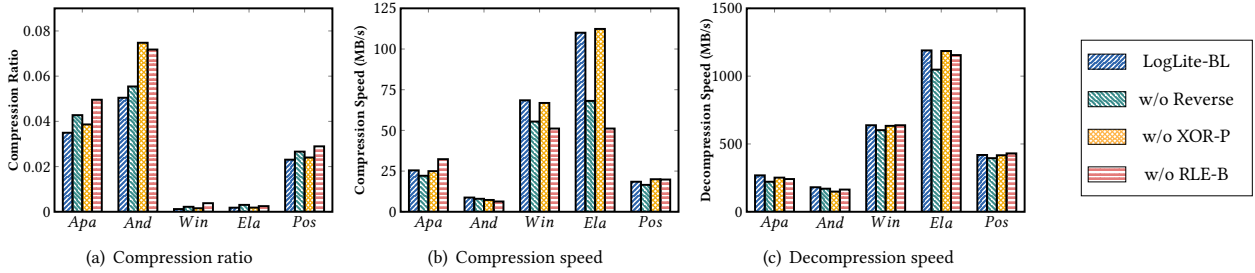
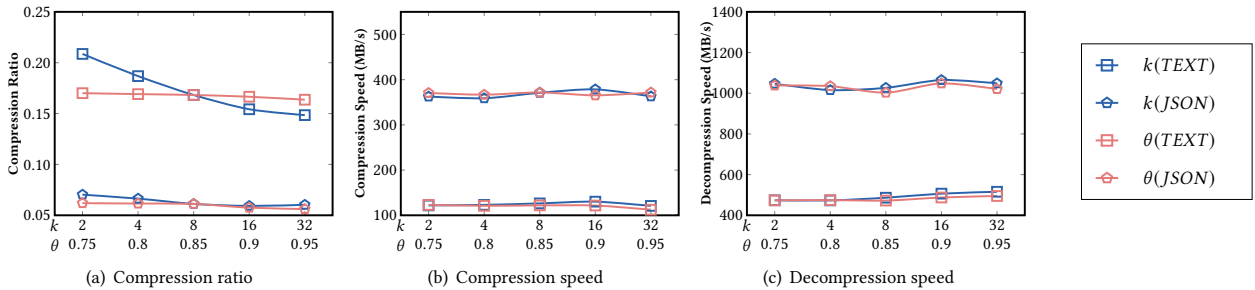
\begin{figure*}
    \centering
    \subfigure[Compression ratio]{
        \begin{tikzpicture}[scale=0.37]
            \begin{axis}[
                width=0.65\textwidth,
                height=0.5\textwidth,
                ymin=0.05,ymax=0.25,
                ytick = {0.05, 0.10, 0.15,0.20,0.25},
                mark size=4.0pt,
                ticklabel style={align=center,font=\Huge},
                xlabel style={font=\LARGE},
                xtick = {0.5,1,2,3,4,5},
                xticklabels={$k$\\$\theta$,$2$\\$0.75$, $4$\\$0.8$, $8$\\$0.85$, $16$\\$0.9$,$32$\\$0.95$},
                yticklabels={0.05, 0.10, 0.15,0.20,0.25},
                xmin=0.5,xmax=5.5,
                ylabel={\huge \bf Compression Ratio},
                ylabel style={yshift=0.2cm},
                every axis plot/.append style={line width = 2pt},
                every axis/.append style={line width = 2pt},
                ]
                \addplot[smooth,mark=square,color=c1,mark size=5pt] plot coordinates {
                    (1,0.208638333333333 )
                    (2,0.186812333333333 )
                    (3,0.16828 )
                    (4,0.154263666666667)
                    (5,0.148484333333333)
                };
                \addplot[smooth,mark=pentagon,color=c1,mark size=5pt] plot coordinates {
                    (1,0.07029725 )
                    (2,0.06638875 )
                    (3,0.06119185 )
                    (4,0.0592088)
                    (5,0.0603352)
                };

                \addplot[smooth,mark=square,color=c5,mark size=5pt] plot coordinates {
                    (1,0.170074666666667 )
                    (2,0.169136333333333 )
                    (3,0.16828 )
                    (4,0.166554)
                    (5,0.163639666666667)
                };
                \addplot[smooth,mark=pentagon,color=c5,mark size=5pt] plot coordinates {
                    (1,0.06194895 )
                    (2,0.0615342 )
                    (3,0.06119185 )
                    (4,0.0575052)
                    (5,0.0561296)
                };
            \end{axis}
        \end{tikzpicture}
    }
    \subfigure[Compression speed]{
        \begin{tikzpicture}[scale=0.37]
            \begin{axis}[
                width=0.65\textwidth,
                height=0.5\textwidth,
                ymin=100,ymax=550,
                ytick = {100, 200, 300,400,500},
                yticklabels={100, 200, 300,400,500},
                mark size=4.0pt,
                ticklabel style={align=center,font=\Huge},
                xlabel style={font=\LARGE},
                xtick = {0.5,1,2,3,4,5},
                xticklabels={$k$\\$\theta$,$2$\\$0.75$, $4$\\$0.8$, $8$\\$0.85$, $16$\\$0.9$,$32$\\$0.95$},
                xmin=0.5,xmax=5.5,
                ylabel={\huge \bf Compression Speed (MB/s)},
                ylabel style={yshift=0.2cm},
                every axis plot/.append style={line width = 2pt},
                every axis/.append style={line width = 2pt},
                ]
                \addplot[smooth,mark=square,color=c1,mark size=5pt] plot coordinates {
                    (1,121.856533333333 )
                    (2,123.033266666667 )
                    (3,126.356633333333 )
                    (4,130.2006)
                    (5,120.799866666667)
                };
                \addplot[smooth,mark=pentagon,color=c1,mark size=5pt] plot coordinates {
                    (1,362.6845)
                    (2,358.904)
                    (3,370.958)
                    (4,378.5445)
                    (5,363.0285)
                };

                \addplot[smooth,mark=square,color=c5,mark size=5pt] plot coordinates {
                    (1,123.009766666667)
                    (2,121.037366666667)
                    (3,121.909666666667)
                    (4,121.609466666667)
                    (5,112.3856)
                };
                \addplot[smooth,mark=pentagon,color=c5,mark size=5pt] plot coordinates {
                    (1,370.609)
                    (2,366.777)
                    (3,371.7685)
                    (4,365.3165)
                    (5,371.6465)
                };
            \end{axis}
        \end{tikzpicture}
    }
    \subfigure[Decompression speed]{
        \begin{tikzpicture}[scale=0.37]
            \begin{axis}[
                width=0.65\textwidth,
                height=0.5\textwidth,
                ymin=400,ymax=1400,
                ytick = {400,600,800,1000,1200,1400},
                yticklabels={400,600,800,1000,1200,1400},
                mark size=4.0pt,
                ticklabel style={align=center,font=\Huge},
                xlabel style={font=\LARGE},
                xtick = {0.5,1,2,3,4,5},
                xticklabels={$k$\\$\theta$,$2$\\$0.75$, $4$\\$0.8$, $8$\\$0.85$, $16$\\$0.9$,$32$\\$0.95$},
                xmin=0.5,xmax=5.5,
                ylabel={\huge \bf Decompression Speed (MB/s)},
                ylabel style={yshift=0.2cm},
                every axis plot/.append style={line width = 2pt},
                every axis/.append style={line width = 2pt},
                ]
                \addplot[smooth,mark=square,color=c1,mark size=5pt] plot coordinates {
                    (1,473.865333333333 )
                    (2,473.092666666667 )
                    (3,486.469333333333 )
                    (4,506.48)
                    (5,515.701666666667)
                };
                \addplot[smooth,mark=pentagon,color=c1,mark size=5pt] plot coordinates {
                    (1,1046.581)
                    (2,1015.4885)
                    (3,1026.4565)
                    (4,1064.915)
                    (5,1048.599)
                };

                \addplot[smooth,mark=square,color=c5,mark size=5pt] plot coordinates {
                    (1,473.617)
                    (2,474.559666666667)
                    (3,472.415)
                    (4,487.464333333333)
                    (5,494.856666666667)
                };
                \addplot[smooth,mark=pentagon,color=c5,mark size=5pt] plot coordinates {
                    (1,1038.128)
                    (2,1034.8145)
                    (3,1002.893)
                    (4,1048.299)
                    (5,1020.508)
                };
            \end{axis}
        \end{tikzpicture}%
    }
    \hspace*{0.5cm} 
    \raisebox{0.72cm}{ 
        \begin{tikzpicture}[scale=0.5]
            \begin{axis}[
                axis line style=none, 
                scale only axis,
                height=0pt, 
                width=0pt,  
                xmin=0, xmax=1, ymin=0, ymax=1, 
                legend style={
                    at={(0,1)}, 
                    anchor=north, 
                    legend columns=1, 
                    minimum width=3.15cm, 
                    minimum height=1cm, 
                    column sep=0.1em, 
                    draw=black, 
                    legend cell align={left}, 
                    inner xsep=10pt, 
                    inner ysep=2pt 
                },
            ]
                \addlegendimage{color=c1, mark=square, smooth, mark size=4pt,line width=1.5pt}
                \addlegendentry{\huge $k(TEXT)$}
                
                \addlegendimage{color=c1, mark=pentagon, smooth, mark size=4pt,line width=1.5pt}
                \addlegendentry{\huge $k(JSON)$}
                
                \addlegendimage{color=c5, mark=square, smooth, mark size=4pt,line width=1.5pt}
                \addlegendentry{\huge $\theta(TEXT)$}

                \addlegendimage{color=c5, mark=pentagon, smooth, mark size=4pt,line width=1.5pt}
                \addlegendentry{\huge $\theta(JSON)$}
                
            \end{axis}
        \end{tikzpicture}
    }
    \vspace{-3mm}
    \caption{Effect of different window size $k$ and similarity threshold $\theta$.}
    \label{fig:parameter}
    \vspace{-1em}
\end{figure*}

\subsection{Case Study}
A key application of \MYLOGNAME{ }is compressing logs line-by-line immediately after they are generated, reducing the resources needed for writing and sending logs. 
We integrate \MYLOGNAME{}-b, PBC, and Zstd-d into PostgreSQL with configuration of \texttt{log\_statement=all} and \texttt{log\_min\_messages=debug5} to enable the full logging strategy.
During stress testing, the database generates a large volume of logs. 
These logs are compressed in real-time by a dedicated thread running the compression algorithm before being written to disk. 
We measure both the compression ratios and the throughput of logs from generation to their compressed output on disk.

The overall PostgreSQL log data compression ratios for \MYLOGNAME{}-b, PBC, and Zstd-d are \textbf{0.1258}, \textbf{0.5802}, and \textbf{0.3255}, respectively. 
Notably, \MYLOGNAME{}-b is plug-and-play, whereas PBC and Zstd-d require offline sampling and training of their respective dictionaries using pre-existing log data.
Due to the dynamic nature of logs and the dictionary of PBC being trained on historical log data, PBC struggles to fully adapt to the log characteristics during compression, resulting in poor compression ratios in this scenario.

As shown in Figure \ref{fig:casestudy}, when the TPS is low, indicating a low log generation rate, all compression algorithms can handle the logs in real time, with throughput increasing as the TPS rises. 
However, when the system has to process logs generated by \num{10e4} transaction/s, both PBC and Zstd-d reach their performance bottlenecks, causing their throughput to plateau. 
In contrast, our method continues to work effectively. 
\MYLOGNAME{}-b achieves a maximum throughput of 2.24 million rows per second, outperforming PBC and Zstd-d by 96.98\% and 63.93\%, respectively.


Theoretically, as long as disk write speed exceeds log generation rate, uncompressed log throughput will increase with log generation rate (i.e., Uncompressed-unlimited). 
However, in real-world environments, hardware I/O limitations or resource constraints cause throughput to hit a bottleneck. 
In the Uncompressed-limited case, we cap the write speed at 250 MB/s, corresponding to typical HDD speeds. 
When processing logs from 150,000 TPS, disk write speed becomes the bottleneck. 
However, \MYLOGNAME{}-b compresses logs, requiring less than 50 MB/s of I/O, resulting in a 23.99\% throughput improvement over Uncompressed-limited.



\subsection{Ablation Study}
To evaluate each component's effectiveness in \MYLOGNAME{}'s design, we conduct an ablation study on \MYLOGNAME{}-BL and its variants: without reverse lookup in L-Windows (w/o Reverse), without preserving original characters in XOR-P (w/o XOR-P), and without byte alignment in RLE-B (w/o RLE-B). 
We select 5 representative datasets, including 3 TEXT datasets (Apache, Android, Windows) and 2 JSON datasets (Elasticsearch, PostgreSQL). 
Compression ratio, compression speed, and decompression speed are presented in Figure \ref{fig:ablation}.

Compared to w/o Reverse, the full \MYLOGNAME{}-BL achieves an average improvement of 14.23\% in compression ratio. 
This is because, although w/o Reverse identifies logs that exceed the default similarity threshold \( \theta \) (i.e., 0.85), the logs found without reverse lookup have lower similarity scores in general. 
In terms of compression speed, the full \MYLOGNAME{}-BL shows an average improvement of 35.83\%, as reverse lookup reduces the number of searches needed to find similar logs. 
For decompression speed, the full \MYLOGNAME{}-BL demonstrates an average improvement of 10.62\%, as the inferior compression ratio in w/o Reverse requires more iterations to decode RLE and XOR.

Compared to w/o XOR-P, the full \MYLOGNAME{}-BL achieves an average improvement of 20.92\% in compression ratio. 
This is because w/o XOR-P applies XOR to all original characters, disrupting the character distribution in logs and making it harder for LZMA to achieve further compression. 
While preserving original characters requires additional operations, the use of efficient SIMD implementations limits the average slowdown in compression speed to just 0.13\%. 
Furthermore, the full \MYLOGNAME{}-BL achieves an average improvement of 2.23\% in decompression speed, as its XOR-P design eliminates the need for XOR computation during decompression.

Compared to w/o RLE-B, the full \MYLOGNAME{}-BL achieves an average improvement of 28.75\% in compression ratio, along with average improvements of 11.14\% in compression speed and 2.55\% in decompression speed, which is because LZMA achieves better compression on byte-aligned streams compared to bit-aligned streams.

\subsection{The Impact of Different Parameter Settings}

This subsection examines the impact of key parameters (window size $k$ and similarity threshold $\theta$) on \MYLOGNAME{}'s performance. We evaluate \MYLOGNAME{}-B with $k$ values of 2, 4, 8, 16, and 32, and $\theta$ values of 0.75, 0.80, 0.85, 0.90, and 0.95. 
Figure \ref{fig:parameter} shows the average changes in compression ratio, compression speed, and decompression speed in the above 3 TEXT and 2 JSON datasets.


Increasing $k$ improves the compression ratio, with a 28.81\% gain for TEXT datasets and 16.58\% for JSON datasets by finding more similar logs in larger windows requiring more memory.
A stricter threshold $\theta$ yields a slight improvement: 3.82\% for TEXT datasets and 9.37\% for JSON datasets on average. 
The improvement varies by dataset, for example, the compression ratio increases by 10.89\% on Windows when $\theta$ rises from 0.75 to 0.95, but only 0.52\% on Elasticsearch. 
Compression and decompression speeds are minimally affected by changes in $k$ and $\theta$. 
While larger $k$ or $\theta$ could increase search time and reduce compression speed, their improvement in compression ratio reduces encoding time, keeping the speed stable. 
Decompression speed tends to increase with larger $k$ and $\theta$, as the improved compression ratio reduces decoding time. In summary, larger $k$ and $\theta$ typically improve compression ratio with minimal speed impact but require more memory and CPU.

In summary, larger $k$ and $\theta$ typically improve compression ratio with minimal speed impact, but require more memory and CPU. Use smaller values in real-time, resource-constrained settings, and larger ones when prioritizing compression efficiency with adequate resources. Different log datasets show varying sensitivity to parameter adjustments, and application requirements differ substantially across use cases. In practice, optimal parameter selection should consider four key factors: log data characteristics, application-specific requirements, desired compression performance metrics, and available computational resources.

\section{Related Work}
\label{sec:rw}
In this section, we introduce the existing works on data compression by categorizing them into three primary areas: General-Purpose Compression, Log Data Compression, and Other Compression.

\subsection{General-Purpose Compression}
General-purpose compression algorithms are the most widely used, as they can effectively compress data without relying on specific data distributions or requiring prior knowledge about the data. 

\textbf{Dictionary- and Statistics-based Compression.} 
Compression algorithms like LZ77~\cite{ziv1977lz77}, LZ78~\cite{ziv1978lz78}, and their various derivatives, such as LZ4~\cite{lz4}, LZMA~\cite{lzma}, Zstd~\cite{collet2018zstd}, DEFLATE~\cite{Deflate,zip,gailly1992gnu}, Snappy~\cite{snappy}, and Brotli~\cite{alakuijala2018brotli}, combine a dictionary matching phase with a fast entropy coding phase. 
The primary idea is to treat data as a byte stream within a sliding window, identify similar byte sequences in this window, and replace repeated bytes with shorter dictionary indices. 
This is followed by further compression using entropy coding techniques such as Huffman coding~\cite{huffmancoding}, Arithmetic coding~\cite{Arithmetic_coding}, and Asymmetric Numeral Systems (ANS)~\cite{ans}, which rely on statistical analysis of the data to assign shorter codes to more frequent symbols, thereby improving compression performance. 
FSST~\cite{boncz2020fsst} encodes strings using a symbol table, which maps 1-8 byte sequences onto single-byte codes.

General compression algorithms vary in speed and compression ratio by adopting different strategies. 
LZ4~\cite{lz4} avoids entropy coding and simplifies dictionary matching, achieving exceptional speed, making it a top choice for general compression~\cite{boncz2020fsst,lzbench} and widely used in Hadoop~\cite{hadoop}. 
LZMA~\cite{lzma} uses a large window, complex context models, and near-optimal arithmetic coding to maximize compression ratio at the cost of speed, making it suitable for cold data archiving. Zstd~\cite{collet2018zstd} balances speed and ratio, reaching the Pareto frontier and being adopted in AWS Redshift~\cite{redshift} and RocksDB~\cite{rocksdb}, and supports offline-trained dictionaries for small data. 


\subsection{Log Data Compression}
Compared to general compression, log data compression takes full advantage of the semantic information of logs, particularly the structured components and the significant similarities within log records, to achieve better compression performance. 
However, most of these methods require accurate and appropriate prior knowledge of the log data, provided either manually or automatically.

\textbf{Non-Parsing-based Compression.} 
LogArchive~\cite{logarchive} adaptively assigns log entries to different buckets using a similarity function to minimize heterogeneity within each bucket and compresses these buckets separately in parallel. 
Cowic~\cite{cowic} employs a semi-static dictionary model and Huffman coding to split log entries into multiple columns, building different models for each column for compression. 
MLC~\cite{feng2016mlc} compression technology identifies and organizes similar log records using Content-Defined Chunking (CDC) and a Jaccard distance-based bucket allocation strategy. It then encodes new log entries using delta compression and ultimately applies general-purpose compression for final compression. 

\textbf{Parsing-based Compression.} 
Since logs typically follow certain patterns or templates, to fully leverage the semantic information of the fields in the logs and achieve higher compression ratios, parsing-based compression methods parse logs and extract patterns or templates using log parsers or regular expressions, albeit at the cost of efficiency, and stores shorter representations instead of the original and repeated patterns or templates. 
LogZip~\cite{liu2019logzip} parses templates from log samples and extracts all templates from raw log files through an iterative matching and extraction process, compressing logs into template IDs and variables. 
LogReducer~\cite{wei2021logreducer} builds on LogZip by analyzing and optimizing it and applies specific encoding techniques for timestamps and numerical variables to achieve better compression ratios. In order to accelerate the template matching, LogReducer segments templates into tokens using delimiters and groups templates with identical token counts during parse tree construction, enabling rapid filtering of candidate templates based on an incoming log entry's token count. Building on LogReducer's parser, LogGrep~\cite{wei2023loggrep} divides logs into subcomponents by identifying common patterns, even splitting variables for finer granularity, and uses tables to map and reconstruct original log messages for more efficient compression and search. LogGrep also leverages the observation that variable parts of the same runtime pattern often exhibit similar lengths to filter and accelerate searches on compressed logs. 
LogShrink~\cite{li2024logshrink} also adopts the parser of LogReducer and leverages a commonality and variability analyzer to identify repetitive patterns and variations in log data, and store similar data in a columnar format by separating fields, leading to high compression ratios. 
CLP~\cite{rodrigues2021clp} splits logs into three components (i.e. timestamps, static text, and variable values) and employs a dictionary to store and index repeated information, thus compressing logs and optimizing search efficiency. 
Unlike these methods for unstructured TEXT logs, \(\mu\)Slope~\cite{wang2024muslope} targets semi-structured JSON logs by merging parse trees to store schema structures, grouping logs with the same schema into structured tables for columnar storage. It uses dictionaries and type-specific encoding to reduce data redundancy, achieving efficient compression and fast searches. 
To save more storage resources during archiving, these compression methods often apply further compression using general-purpose compression like Zstd or LZMA.

\subsection{Other Compression}
For the compression of floating-point sequences, since a floating-point number is typically fixed at 32 or 64 bits and many floating-point sequences generated by sensors are nearly continuous, Gorilla~\cite{pelkonen2015gorilla}, Chimp~\cite{liakos2022chimp}, and Elf~\cite{li2023elf} use XOR with the previous value to achieve lightweight compression.
To bridge the gap between general-purpose compression and specific-purpose compression, PBC~\cite{zhang2023high} supports data with patterns such as machine-generated KV data, logs, JSON, URLs, and UUIDs, etc. 
PBC extracts potential patterns from the data through clustering, thereby reducing redundancy in common subsequences, and achieves a high compression ratio, fast decompression speed, and competitive compression speed without any manually defined knowledge.



\section{Conclusion}
\label{sec:conclusion}

We present \MYLOGNAME{}, a lightweight plug-and-play streaming lossless compression algorithm for TEXT and JSON logs. 
Based on our in-depth characterization study, it exploits the strong correlation between log length and similarity. 
Our evaluation demonstrates that \MYLOGNAME{ }achieves Pareto-optimal performance in most scenarios.



\begin{acks}
This work is supported by the National Key R\&D Program of China (2022YFB3103700), CCF-AFSG Research Fund (RF20220217), ARC Future Fellowship (FT210100303), ARC Discovery Project (DP230101445), Guangdong S\&T Program under Grant 2024B01010\\10002, National Natural Science Foundation of China (No.U2436208, No.62372129), and Project of Guangdong Key Laboratory of Industrial Control System Security (2024B1212020010).
\end{acks}


\bibliographystyle{ACM-Reference-Format}
\bibliography{compression}


\begin{thebibliography}{60}


\ifx \showCODEN    \undefined \def \showCODEN     #1{\unskip}     \fi
\ifx \showDOI      \undefined \def \showDOI       #1{#1}\fi
\ifx \showISBNx    \undefined \def \showISBNx     #1{\unskip}     \fi
\ifx \showISBNxiii \undefined \def \showISBNxiii  #1{\unskip}     \fi
\ifx \showISSN     \undefined \def \showISSN      #1{\unskip}     \fi
\ifx \showLCCN     \undefined \def \showLCCN      #1{\unskip}     \fi
\ifx \shownote     \undefined \def \shownote      #1{#1}          \fi
\ifx \showarticletitle \undefined \def \showarticletitle #1{#1}   \fi
\ifx \showURL      \undefined \def \showURL       {\relax}        \fi
\providecommand\bibfield[2]{#2}
\providecommand\bibinfo[2]{#2}
\providecommand\natexlab[1]{#1}
\providecommand\showeprint[2][]{arXiv:#2}

\bibitem[\protect\citeauthoryear{??}{log}{2025}]%
        {loglitecode}
 \bibinfo{year}{2025}\natexlab{}.
\newblock \bibinfo{title}{LogLite}.
\newblock
\newblock
\urldef\tempurl%
\url{https://github.com/benzhaotang/LogLite}
\showURL{%
\tempurl}


\bibitem[\protect\citeauthoryear{Agrawal, Kotia, Moshirian, and Kim}{Agrawal et~al\mbox{.}}{2018}]%
        {agrawal2018log}
\bibfield{author}{\bibinfo{person}{Vaibhav Agrawal}, \bibinfo{person}{Devanjal Kotia}, \bibinfo{person}{Kamelia Moshirian}, {and} \bibinfo{person}{Mihui Kim}.} \bibinfo{year}{2018}\natexlab{}.
\newblock \showarticletitle{Log-based cloud monitoring system for OpenStack}. In \bibinfo{booktitle}{\emph{2018 IEEE Fourth International Conference on Big Data Computing Service and Applications (BigDataService)}}. IEEE, \bibinfo{pages}{276--281}.
\newblock


\bibitem[\protect\citeauthoryear{Alakuijala, Farruggia, Ferragina, Kliuchnikov, Obryk, Szabadka, and Vandevenne}{Alakuijala et~al\mbox{.}}{2018}]%
        {alakuijala2018brotli}
\bibfield{author}{\bibinfo{person}{Jyrki Alakuijala}, \bibinfo{person}{Andrea Farruggia}, \bibinfo{person}{Paolo Ferragina}, \bibinfo{person}{Eugene Kliuchnikov}, \bibinfo{person}{Robert Obryk}, \bibinfo{person}{Zoltan Szabadka}, {and} \bibinfo{person}{Lode Vandevenne}.} \bibinfo{year}{2018}\natexlab{}.
\newblock \showarticletitle{Brotli: A general-purpose data compressor}.
\newblock \bibinfo{journal}{\emph{ACM Transactions on Information Systems (TOIS)}} \bibinfo{volume}{37}, \bibinfo{number}{1} (\bibinfo{year}{2018}), \bibinfo{pages}{1--30}.
\newblock


\bibitem[\protect\citeauthoryear{aliyun}{aliyun}{2018}]%
        {aliyunIoT}
\bibfield{author}{\bibinfo{person}{aliyun}.} \bibinfo{year}{2018}\natexlab{}.
\newblock \bibinfo{title}{aliyunIoT.}
\newblock
\newblock
\urldef\tempurl%
\url{https://developer.aliyun.com/article/637406}
\showURL{%
\tempurl}


\bibitem[\protect\citeauthoryear{Amar and Rigby}{Amar and Rigby}{2019}]%
        {amar2019mining}
\bibfield{author}{\bibinfo{person}{Anunay Amar} {and} \bibinfo{person}{Peter~C Rigby}.} \bibinfo{year}{2019}\natexlab{}.
\newblock \showarticletitle{Mining historical test logs to predict bugs and localize faults in the test logs}. In \bibinfo{booktitle}{\emph{2019 IEEE/ACM 41st International Conference on Software Engineering (ICSE)}}. IEEE, \bibinfo{pages}{140--151}.
\newblock


\bibitem[\protect\citeauthoryear{Boncz, Neumann, and Leis}{Boncz et~al\mbox{.}}{2020}]%
        {boncz2020fsst}
\bibfield{author}{\bibinfo{person}{Peter Boncz}, \bibinfo{person}{Thomas Neumann}, {and} \bibinfo{person}{Viktor Leis}.} \bibinfo{year}{2020}\natexlab{}.
\newblock \showarticletitle{FSST: fast random access string compression}.
\newblock \bibinfo{journal}{\emph{Proceedings of the VLDB Endowment}} \bibinfo{volume}{13}, \bibinfo{number}{12} (\bibinfo{year}{2020}), \bibinfo{pages}{2649--2661}.
\newblock


\bibitem[\protect\citeauthoryear{Cao, Feng, Liang, Zhang, Gao, Zhang, and Li}{Cao et~al\mbox{.}}{2021}]%
        {cao2021logstore}
\bibfield{author}{\bibinfo{person}{Wei Cao}, \bibinfo{person}{Xiaojie Feng}, \bibinfo{person}{Boyuan Liang}, \bibinfo{person}{Tianyu Zhang}, \bibinfo{person}{Yusong Gao}, \bibinfo{person}{Yunyang Zhang}, {and} \bibinfo{person}{Feifei Li}.} \bibinfo{year}{2021}\natexlab{}.
\newblock \showarticletitle{Logstore: A cloud-native and multi-tenant log database}. In \bibinfo{booktitle}{\emph{Proceedings of the 2021 International Conference on Management of Data}}. \bibinfo{pages}{2464--2476}.
\newblock


\bibitem[\protect\citeauthoryear{Christensen and Li}{Christensen and Li}{2013}]%
        {logarchive}
\bibfield{author}{\bibinfo{person}{Robert Christensen} {and} \bibinfo{person}{Feifei Li}.} \bibinfo{year}{2013}\natexlab{}.
\newblock \showarticletitle{Adaptive log compression for massive log data.}. In \bibinfo{booktitle}{\emph{SIGMOD Conference}}. \bibinfo{pages}{1283--1284}.
\newblock


\bibitem[\protect\citeauthoryear{Collet}{Collet}{2011}]%
        {lz4}
\bibfield{author}{\bibinfo{person}{Yann Collet}.} \bibinfo{year}{2011}\natexlab{}.
\newblock \bibinfo{title}{LZ4: Fast Compression Algorithm}.
\newblock
\newblock
\urldef\tempurl%
\url{https://github.com/lz4/lz4}
\showURL{%
\tempurl}


\bibitem[\protect\citeauthoryear{Collet and Kucherawy}{Collet and Kucherawy}{2018}]%
        {collet2018zstd}
\bibfield{author}{\bibinfo{person}{Yann Collet} {and} \bibinfo{person}{Murray Kucherawy}.} \bibinfo{year}{2018}\natexlab{}.
\newblock \bibinfo{booktitle}{\emph{Zstandard Compression and the application/zstd Media Type}}.
\newblock \bibinfo{type}{{T}echnical {R}eport}.
\newblock


\bibitem[\protect\citeauthoryear{Du, Li, Zheng, and Srikumar}{Du et~al\mbox{.}}{2017}]%
        {du2017deeplog}
\bibfield{author}{\bibinfo{person}{Min Du}, \bibinfo{person}{Feifei Li}, \bibinfo{person}{Guineng Zheng}, {and} \bibinfo{person}{Vivek Srikumar}.} \bibinfo{year}{2017}\natexlab{}.
\newblock \showarticletitle{Deeplog: Anomaly detection and diagnosis from system logs through deep learning}. In \bibinfo{booktitle}{\emph{Proceedings of the 2017 ACM SIGSAC conference on computer and communications security}}. \bibinfo{pages}{1285--1298}.
\newblock


\bibitem[\protect\citeauthoryear{Duda, Tahboub, Gadgil, and Delp}{Duda et~al\mbox{.}}{2015}]%
        {ans}
\bibfield{author}{\bibinfo{person}{Jarek Duda}, \bibinfo{person}{Khalid Tahboub}, \bibinfo{person}{Neeraj~J Gadgil}, {and} \bibinfo{person}{Edward~J Delp}.} \bibinfo{year}{2015}\natexlab{}.
\newblock \showarticletitle{The use of asymmetric numeral systems as an accurate replacement for Huffman coding}. In \bibinfo{booktitle}{\emph{2015 Picture Coding Symposium (PCS)}}. IEEE, \bibinfo{pages}{65--69}.
\newblock


\bibitem[\protect\citeauthoryear{Dumais, Jeffries, Russell, Tang, and Teevan}{Dumais et~al\mbox{.}}{2014}]%
        {dumais2014understanding}
\bibfield{author}{\bibinfo{person}{Susan Dumais}, \bibinfo{person}{Robin Jeffries}, \bibinfo{person}{Daniel~M Russell}, \bibinfo{person}{Diane Tang}, {and} \bibinfo{person}{Jaime Teevan}.} \bibinfo{year}{2014}\natexlab{}.
\newblock \showarticletitle{Understanding user behavior through log data and analysis}.
\newblock \bibinfo{journal}{\emph{Ways of Knowing in HCI}} (\bibinfo{year}{2014}), \bibinfo{pages}{349--372}.
\newblock


\bibitem[\protect\citeauthoryear{Fan, Chen, Tung, Wu, and Chen}{Fan et~al\mbox{.}}{2015}]%
        {fan2015framework}
\bibfield{author}{\bibinfo{person}{Yao-Chung Fan}, \bibinfo{person}{Yu-Chi Chen}, \bibinfo{person}{Kuan-Chieh Tung}, \bibinfo{person}{Kuo-Chen Wu}, {and} \bibinfo{person}{Arbee~LP Chen}.} \bibinfo{year}{2015}\natexlab{}.
\newblock \showarticletitle{A framework for enabling user preference profiling through wi-fi logs}.
\newblock \bibinfo{journal}{\emph{IEEE Transactions on Knowledge and Data Engineering}} \bibinfo{volume}{28}, \bibinfo{number}{3} (\bibinfo{year}{2015}), \bibinfo{pages}{592--603}.
\newblock


\bibitem[\protect\citeauthoryear{Fazzinga, Flesca, Furfaro, and Pontieri}{Fazzinga et~al\mbox{.}}{2018}]%
        {fazzinga2018online}
\bibfield{author}{\bibinfo{person}{Bettina Fazzinga}, \bibinfo{person}{Sergio Flesca}, \bibinfo{person}{Filippo Furfaro}, {and} \bibinfo{person}{Luigi Pontieri}.} \bibinfo{year}{2018}\natexlab{}.
\newblock \showarticletitle{Online and offline classification of traces of event logs on the basis of security risks}.
\newblock \bibinfo{journal}{\emph{Journal of Intelligent Information Systems}}  \bibinfo{volume}{50} (\bibinfo{year}{2018}), \bibinfo{pages}{195--230}.
\newblock


\bibitem[\protect\citeauthoryear{Feng, Wu, and Li}{Feng et~al\mbox{.}}{2016}]%
        {feng2016mlc}
\bibfield{author}{\bibinfo{person}{Bo Feng}, \bibinfo{person}{Chentao Wu}, {and} \bibinfo{person}{Jie Li}.} \bibinfo{year}{2016}\natexlab{}.
\newblock \showarticletitle{MLC: an efficient multi-level log compression method for cloud backup systems}. In \bibinfo{booktitle}{\emph{2016 IEEE Trustcom/BigDataSE/ISPA}}. IEEE, \bibinfo{pages}{1358--1365}.
\newblock


\bibitem[\protect\citeauthoryear{Gailly and Adler}{Gailly and Adler}{1992}]%
        {gailly1992gnu}
\bibfield{author}{\bibinfo{person}{Jean-loup Gailly} {and} \bibinfo{person}{Mark Adler}.} \bibinfo{year}{1992}\natexlab{}.
\newblock \showarticletitle{GNU gzip}.
\newblock \bibinfo{journal}{\emph{GNU Operating System}} (\bibinfo{year}{1992}).
\newblock


\bibitem[\protect\citeauthoryear{He, Lin, Lou, Zhang, Lyu, and Zhang}{He et~al\mbox{.}}{2018}]%
        {he2018identifying}
\bibfield{author}{\bibinfo{person}{Shilin He}, \bibinfo{person}{Qingwei Lin}, \bibinfo{person}{Jian-Guang Lou}, \bibinfo{person}{Hongyu Zhang}, \bibinfo{person}{Michael~R Lyu}, {and} \bibinfo{person}{Dongmei Zhang}.} \bibinfo{year}{2018}\natexlab{}.
\newblock \showarticletitle{Identifying impactful service system problems via log analysis}. In \bibinfo{booktitle}{\emph{Proceedings of the 2018 26th ACM joint meeting on European software engineering conference and symposium on the foundations of software engineering}}. \bibinfo{pages}{60--70}.
\newblock


\bibitem[\protect\citeauthoryear{Huang, Huang, Dai, Xie, and Huang}{Huang et~al\mbox{.}}{2010}]%
        {huang2010hibench}
\bibfield{author}{\bibinfo{person}{Shengsheng Huang}, \bibinfo{person}{Jie Huang}, \bibinfo{person}{Jinquan Dai}, \bibinfo{person}{Tao Xie}, {and} \bibinfo{person}{Bo Huang}.} \bibinfo{year}{2010}\natexlab{}.
\newblock \showarticletitle{The HiBench benchmark suite: Characterization of the MapReduce-based data analysis}. In \bibinfo{booktitle}{\emph{2010 IEEE 26th International conference on data engineering workshops (ICDEW 2010)}}. IEEE, \bibinfo{pages}{41--51}.
\newblock


\bibitem[\protect\citeauthoryear{Huffman}{Huffman}{1952}]%
        {huffmancoding}
\bibfield{author}{\bibinfo{person}{David~A. Huffman}.} \bibinfo{year}{1952}\natexlab{}.
\newblock \showarticletitle{A Method for the Construction of Minimum-Redundancy Codes}.
\newblock \bibinfo{journal}{\emph{Proceedings of the IRE}} \bibinfo{volume}{40}, \bibinfo{number}{9} (\bibinfo{year}{1952}), \bibinfo{pages}{1098--1101}.
\newblock


\bibitem[\protect\citeauthoryear{Inc.}{Inc.}{2012}]%
        {rocksdb}
\bibfield{author}{\bibinfo{person}{Facebook Inc.}} \bibinfo{year}{2012}\natexlab{}.
\newblock \bibinfo{title}{RocksDB: A Persistent Key-Value Store for Flash and RAM Storage}.
\newblock
\newblock
\urldef\tempurl%
\url{https://github.com/facebook/rocksdb}
\showURL{%
\tempurl}


\bibitem[\protect\citeauthoryear{Inc.}{Inc.}{2011}]%
        {snappy}
\bibfield{author}{\bibinfo{person}{Google Inc.}} \bibinfo{year}{2011}\natexlab{}.
\newblock \bibinfo{title}{Snappy: A Fast Compressor}.
\newblock
\newblock
\urldef\tempurl%
\url{https://github.com/google/snappy}
\showURL{%
\tempurl}


\bibitem[\protect\citeauthoryear{Industries}{Industries}{2021}]%
        {pareto_opt_compression}
\bibfield{author}{\bibinfo{person}{Insanity Industries}.} \bibinfo{year}{2021}\natexlab{}.
\newblock \bibinfo{title}{Pareto-optimal compression.}
\newblock
\newblock
\urldef\tempurl%
\url{https://insanity.industries/post/pareto-optimal-compression/}
\showURL{%
\tempurl}


\bibitem[\protect\citeauthoryear{Jia, Li, Liu, Liao, and Liu}{Jia et~al\mbox{.}}{2018}]%
        {jia2018smartlog}
\bibfield{author}{\bibinfo{person}{Zhouyang Jia}, \bibinfo{person}{Shanshan Li}, \bibinfo{person}{Xiaodong Liu}, \bibinfo{person}{Xiangke Liao}, {and} \bibinfo{person}{Yunhuai Liu}.} \bibinfo{year}{2018}\natexlab{}.
\newblock \showarticletitle{SMARTLOG: Place error log statement by deep understanding of log intention}. In \bibinfo{booktitle}{\emph{2018 IEEE 25th International Conference on Software Analysis, Evolution and Reengineering (SANER)}}. IEEE, \bibinfo{pages}{61--71}.
\newblock


\bibitem[\protect\citeauthoryear{Katz}{Katz}{1989}]%
        {zip}
\bibfield{author}{\bibinfo{person}{Phil Katz}.} \bibinfo{year}{1989}\natexlab{}.
\newblock \bibinfo{title}{ZIP: A File Compression Standard}.
\newblock
\newblock
\urldef\tempurl%
\url{https://www.info-zip.org/}
\showURL{%
\tempurl}


\bibitem[\protect\citeauthoryear{Lempel and Ziv}{Lempel and Ziv}{2001}]%
        {lzma}
\bibfield{author}{\bibinfo{person}{Abraham Lempel} {and} \bibinfo{person}{Jacob Ziv}.} \bibinfo{year}{2001}\natexlab{}.
\newblock \bibinfo{title}{LZMA Algorithm}.
\newblock
\newblock
\urldef\tempurl%
\url{http://www.7-zip.org/}
\showURL{%
\tempurl}


\bibitem[\protect\citeauthoryear{Leshchinskiy}{Leshchinskiy}{2018}]%
        {lzbench}
\bibfield{author}{\bibinfo{person}{Roman Leshchinskiy}.} \bibinfo{year}{2018}\natexlab{}.
\newblock \bibinfo{title}{LZBench: Compression Benchmarking Tool}.
\newblock
\newblock
\urldef\tempurl%
\url{https://github.com/lemire/LZBench}
\showURL{%
\tempurl}


\bibitem[\protect\citeauthoryear{Li, Li, Wu, Chen, and Zheng}{Li et~al\mbox{.}}{2023}]%
        {li2023elf}
\bibfield{author}{\bibinfo{person}{Ruiyuan Li}, \bibinfo{person}{Zheng Li}, \bibinfo{person}{Yi Wu}, \bibinfo{person}{Chao Chen}, {and} \bibinfo{person}{Yu Zheng}.} \bibinfo{year}{2023}\natexlab{}.
\newblock \showarticletitle{Elf: Erasing-based lossless floating-point compression}.
\newblock \bibinfo{journal}{\emph{Proceedings of the VLDB Endowment}} \bibinfo{volume}{16}, \bibinfo{number}{7} (\bibinfo{year}{2023}), \bibinfo{pages}{1763--1776}.
\newblock


\bibitem[\protect\citeauthoryear{Li, Chen, Jing, He, and Yu}{Li et~al\mbox{.}}{2022}]%
        {li2022swisslog}
\bibfield{author}{\bibinfo{person}{Xiaoyun Li}, \bibinfo{person}{Pengfei Chen}, \bibinfo{person}{Linxiao Jing}, \bibinfo{person}{Zilong He}, {and} \bibinfo{person}{Guangba Yu}.} \bibinfo{year}{2022}\natexlab{}.
\newblock \showarticletitle{Swisslog: Robust anomaly detection and localization for interleaved unstructured logs}.
\newblock \bibinfo{journal}{\emph{IEEE Transactions on Dependable and Secure Computing}} \bibinfo{volume}{20}, \bibinfo{number}{4} (\bibinfo{year}{2022}), \bibinfo{pages}{2762--2780}.
\newblock


\bibitem[\protect\citeauthoryear{Li, Zhang, Le, and Chen}{Li et~al\mbox{.}}{2024}]%
        {li2024logshrink}
\bibfield{author}{\bibinfo{person}{Xiaoyun Li}, \bibinfo{person}{Hongyu Zhang}, \bibinfo{person}{Van-Hoang Le}, {and} \bibinfo{person}{Pengfei Chen}.} \bibinfo{year}{2024}\natexlab{}.
\newblock \showarticletitle{Logshrink: Effective log compression by leveraging commonality and variability of log data}. In \bibinfo{booktitle}{\emph{Proceedings of the 46th IEEE/ACM International Conference on Software Engineering}}. \bibinfo{pages}{1--12}.
\newblock


\bibitem[\protect\citeauthoryear{Liakos, Papakonstantinopoulou, and Kotidis}{Liakos et~al\mbox{.}}{2022}]%
        {liakos2022chimp}
\bibfield{author}{\bibinfo{person}{Panagiotis Liakos}, \bibinfo{person}{Katia Papakonstantinopoulou}, {and} \bibinfo{person}{Yannis Kotidis}.} \bibinfo{year}{2022}\natexlab{}.
\newblock \showarticletitle{Chimp: efficient lossless floating point compression for time series databases}.
\newblock \bibinfo{journal}{\emph{Proceedings of the VLDB Endowment}} \bibinfo{volume}{15}, \bibinfo{number}{11} (\bibinfo{year}{2022}), \bibinfo{pages}{3058--3070}.
\newblock


\bibitem[\protect\citeauthoryear{Lin, Zhou, Yao, Guo, and Li}{Lin et~al\mbox{.}}{2015}]%
        {cowic}
\bibfield{author}{\bibinfo{person}{Hao Lin}, \bibinfo{person}{Jingyu Zhou}, \bibinfo{person}{Bin Yao}, \bibinfo{person}{Minyi Guo}, {and} \bibinfo{person}{Jie Li}.} \bibinfo{year}{2015}\natexlab{}.
\newblock \showarticletitle{Cowic: A column-wise independent compression for log stream analysis}. In \bibinfo{booktitle}{\emph{2015 15th IEEE/ACM International Symposium on Cluster, Cloud and Grid Computing}}. IEEE, \bibinfo{pages}{21--30}.
\newblock


\bibitem[\protect\citeauthoryear{Lin et~al\mbox{.}}{Lin et~al\mbox{.}}{2010}]%
        {ycsb}
\bibfield{author}{\bibinfo{person}{Jimmy Lin} {et~al\mbox{.}}} \bibinfo{year}{2010}\natexlab{}.
\newblock \bibinfo{title}{YCSB: A Workload Generation Framework for Cloud Databases}.
\newblock
\newblock
\urldef\tempurl%
\url{https://github.com/brianfrankcooper/YCSB}
\showURL{%
\tempurl}


\bibitem[\protect\citeauthoryear{Liu, Paparrizos, and Elmore}{Liu et~al\mbox{.}}{2024}]%
        {liu2024adaedge}
\bibfield{author}{\bibinfo{person}{Chunwei Liu}, \bibinfo{person}{John Paparrizos}, {and} \bibinfo{person}{Aaron~J Elmore}.} \bibinfo{year}{2024}\natexlab{}.
\newblock \showarticletitle{Adaedge: A dynamic compression selection framework for resource constrained devices}. In \bibinfo{booktitle}{\emph{2024 IEEE 40th International Conference on Data Engineering (ICDE)}}. IEEE, \bibinfo{pages}{1506--1519}.
\newblock


\bibitem[\protect\citeauthoryear{Liu, Zhu, He, He, Zheng, and Lyu}{Liu et~al\mbox{.}}{2019}]%
        {liu2019logzip}
\bibfield{author}{\bibinfo{person}{Jinyang Liu}, \bibinfo{person}{Jieming Zhu}, \bibinfo{person}{Shilin He}, \bibinfo{person}{Pinjia He}, \bibinfo{person}{Zibin Zheng}, {and} \bibinfo{person}{Michael~R Lyu}.} \bibinfo{year}{2019}\natexlab{}.
\newblock \showarticletitle{Logzip: Extracting hidden structures via iterative clustering for log compression}. In \bibinfo{booktitle}{\emph{2019 34th IEEE/ACM International Conference on Automated Software Engineering (ASE)}}. IEEE, \bibinfo{pages}{863--873}.
\newblock


\bibitem[\protect\citeauthoryear{Mastropaolo, Pascarella, and Bavota}{Mastropaolo et~al\mbox{.}}{2022}]%
        {mastropaolo2022using}
\bibfield{author}{\bibinfo{person}{Antonio Mastropaolo}, \bibinfo{person}{Luca Pascarella}, {and} \bibinfo{person}{Gabriele Bavota}.} \bibinfo{year}{2022}\natexlab{}.
\newblock \showarticletitle{Using deep learning to generate complete log statements}. In \bibinfo{booktitle}{\emph{Proceedings of the 44th International Conference on Software Engineering}}. \bibinfo{pages}{2279--2290}.
\newblock


\bibitem[\protect\citeauthoryear{Meng, Liu, Zhu, Zhang, Pei, Liu, Chen, Zhang, Tao, Sun, et~al\mbox{.}}{Meng et~al\mbox{.}}{2019}]%
        {meng2019loganomaly}
\bibfield{author}{\bibinfo{person}{Weibin Meng}, \bibinfo{person}{Ying Liu}, \bibinfo{person}{Yichen Zhu}, \bibinfo{person}{Shenglin Zhang}, \bibinfo{person}{Dan Pei}, \bibinfo{person}{Yuqing Liu}, \bibinfo{person}{Yihao Chen}, \bibinfo{person}{Ruizhi Zhang}, \bibinfo{person}{Shimin Tao}, \bibinfo{person}{Pei Sun}, {et~al\mbox{.}}} \bibinfo{year}{2019}\natexlab{}.
\newblock \showarticletitle{Loganomaly: Unsupervised detection of sequential and quantitative anomalies in unstructured logs.}. In \bibinfo{booktitle}{\emph{IJCAI}}, Vol.~\bibinfo{volume}{19}. \bibinfo{pages}{4739--4745}.
\newblock


\bibitem[\protect\citeauthoryear{Oprea, Li, Yen, Chin, and Alrwais}{Oprea et~al\mbox{.}}{2015}]%
        {oprea2015detection}
\bibfield{author}{\bibinfo{person}{Alina Oprea}, \bibinfo{person}{Zhou Li}, \bibinfo{person}{Ting-Fang Yen}, \bibinfo{person}{Sang~H Chin}, {and} \bibinfo{person}{Sumayah Alrwais}.} \bibinfo{year}{2015}\natexlab{}.
\newblock \showarticletitle{Detection of early-stage enterprise infection by mining large-scale log data}. In \bibinfo{booktitle}{\emph{2015 45th Annual IEEE/IFIP International Conference on Dependable Systems and Networks}}. IEEE, \bibinfo{pages}{45--56}.
\newblock


\bibitem[\protect\citeauthoryear{Pasco}{Pasco}{1977}]%
        {Arithmetic_coding}
\bibfield{author}{\bibinfo{person}{R. Pasco}.} \bibinfo{year}{1977}\natexlab{}.
\newblock \showarticletitle{Source coding algorithms for fast data compression (Ph.D. Thesis abstr.)}.
\newblock \bibinfo{journal}{\emph{IEEE Transactions on Information Theory}} \bibinfo{volume}{23}, \bibinfo{number}{4} (\bibinfo{year}{1977}), \bibinfo{pages}{548--548}.
\newblock


\bibitem[\protect\citeauthoryear{Pelkonen, Franklin, Teller, Cavallaro, Huang, Meza, and Veeraraghavan}{Pelkonen et~al\mbox{.}}{2015}]%
        {pelkonen2015gorilla}
\bibfield{author}{\bibinfo{person}{Tuomas Pelkonen}, \bibinfo{person}{Scott Franklin}, \bibinfo{person}{Justin Teller}, \bibinfo{person}{Paul Cavallaro}, \bibinfo{person}{Qi Huang}, \bibinfo{person}{Justin Meza}, {and} \bibinfo{person}{Kaushik Veeraraghavan}.} \bibinfo{year}{2015}\natexlab{}.
\newblock \showarticletitle{Gorilla: A fast, scalable, in-memory time series database}.
\newblock \bibinfo{journal}{\emph{Proceedings of the VLDB Endowment}} \bibinfo{volume}{8}, \bibinfo{number}{12} (\bibinfo{year}{2015}), \bibinfo{pages}{1816--1827}.
\newblock


\bibitem[\protect\citeauthoryear{Rodrigues, Luo, and Yuan}{Rodrigues et~al\mbox{.}}{2021}]%
        {rodrigues2021clp}
\bibfield{author}{\bibinfo{person}{Kirk Rodrigues}, \bibinfo{person}{Yu Luo}, {and} \bibinfo{person}{Ding Yuan}.} \bibinfo{year}{2021}\natexlab{}.
\newblock \showarticletitle{CLP: Efficient and scalable search on compressed text logs}. In \bibinfo{booktitle}{\emph{15th USENIX Symposium on Operating Systems Design and Implementation (OSDI 21)}}. \bibinfo{pages}{183--198}.
\newblock


\bibitem[\protect\citeauthoryear{Rong, Gu, Shen, Zhang, and Kuang}{Rong et~al\mbox{.}}{2023}]%
        {rong2023developers}
\bibfield{author}{\bibinfo{person}{Guoping Rong}, \bibinfo{person}{Shenghui Gu}, \bibinfo{person}{Haifeng Shen}, \bibinfo{person}{He Zhang}, {and} \bibinfo{person}{Hongyu Kuang}.} \bibinfo{year}{2023}\natexlab{}.
\newblock \showarticletitle{How Do Developers' Profiles and Experiences Influence their Logging Practices? An Empirical Study of Industrial Practitioners}. In \bibinfo{booktitle}{\emph{2023 IEEE/ACM 45th International Conference on Software Engineering (ICSE)}}. IEEE, \bibinfo{pages}{855--867}.
\newblock


\bibitem[\protect\citeauthoryear{Rosenberg and Moonen}{Rosenberg and Moonen}{2020}]%
        {rosenberg2020spectrum}
\bibfield{author}{\bibinfo{person}{Carl~Martin Rosenberg} {and} \bibinfo{person}{Leon Moonen}.} \bibinfo{year}{2020}\natexlab{}.
\newblock \showarticletitle{Spectrum-based log diagnosis}. In \bibinfo{booktitle}{\emph{Proceedings of the 14th ACM/IEEE International Symposium on Empirical Software Engineering and Measurement (ESEM)}}. \bibinfo{pages}{1--12}.
\newblock


\bibitem[\protect\citeauthoryear{Services}{Services}{2012}]%
        {redshift}
\bibfield{author}{\bibinfo{person}{Amazon~Web Services}.} \bibinfo{year}{2012}\natexlab{}.
\newblock \bibinfo{title}{Amazon Redshift: Data Warehousing Service}.
\newblock
\newblock
\urldef\tempurl%
\url{https://aws.amazon.com/redshift/}
\showURL{%
\tempurl}


\bibitem[\protect\citeauthoryear{Shvachko, Kuang, Radia, and Chansler}{Shvachko et~al\mbox{.}}{2010}]%
        {hadoop}
\bibfield{author}{\bibinfo{person}{Konstantin Shvachko}, \bibinfo{person}{Hairong Kuang}, \bibinfo{person}{Sanjay Radia}, {and} \bibinfo{person}{Robert Chansler}.} \bibinfo{year}{2010}\natexlab{}.
\newblock \showarticletitle{The hadoop distributed file system}. In \bibinfo{booktitle}{\emph{2010 IEEE 26th symposium on mass storage systems and technologies (MSST)}}. Ieee, \bibinfo{pages}{1--10}.
\newblock


\bibitem[\protect\citeauthoryear{Wang, Gibson, Rodrigues, Luo, Zhang, Wang, Fu, Chen, and Yuan}{Wang et~al\mbox{.}}{2024}]%
        {wang2024muslope}
\bibfield{author}{\bibinfo{person}{Rui Wang}, \bibinfo{person}{Devin Gibson}, \bibinfo{person}{Kirk Rodrigues}, \bibinfo{person}{Yu Luo}, \bibinfo{person}{Yun Zhang}, \bibinfo{person}{Kaibo Wang}, \bibinfo{person}{Yupeng Fu}, \bibinfo{person}{Ting Chen}, {and} \bibinfo{person}{Ding Yuan}.} \bibinfo{year}{2024}\natexlab{}.
\newblock \showarticletitle{$\mu$Slope: High Compression and Fast Search on Semi-Structured Logs}. In \bibinfo{booktitle}{\emph{18th USENIX Symposium on Operating Systems Design and Implementation (OSDI 24)}}. \bibinfo{pages}{529--544}.
\newblock


\bibitem[\protect\citeauthoryear{Wang and Chen}{Wang and Chen}{2017}]%
        {wang2017exploiting}
\bibfield{author}{\bibinfo{person}{Zhiyi Wang} {and} \bibinfo{person}{Shimin Chen}.} \bibinfo{year}{2017}\natexlab{}.
\newblock \showarticletitle{Exploiting common patterns for tree-structured data}. In \bibinfo{booktitle}{\emph{Proceedings of the 2017 ACM International Conference on Management of Data}}. \bibinfo{pages}{883--896}.
\newblock


\bibitem[\protect\citeauthoryear{Wei, Zhang, Chen, Wang, Zheng, Sun, Wu, and Jiang}{Wei et~al\mbox{.}}{2023}]%
        {wei2023loggrep}
\bibfield{author}{\bibinfo{person}{Junyu Wei}, \bibinfo{person}{Guangyan Zhang}, \bibinfo{person}{Junchao Chen}, \bibinfo{person}{Yang Wang}, \bibinfo{person}{Weimin Zheng}, \bibinfo{person}{Tingtao Sun}, \bibinfo{person}{Jiesheng Wu}, {and} \bibinfo{person}{Jiangwei Jiang}.} \bibinfo{year}{2023}\natexlab{}.
\newblock \showarticletitle{Loggrep: Fast and cheap cloud log storage by exploiting both static and runtime patterns}. In \bibinfo{booktitle}{\emph{Proceedings of the Eighteenth European Conference on Computer Systems}}. \bibinfo{pages}{452--468}.
\newblock


\bibitem[\protect\citeauthoryear{Wei, Zhang, Wang, Liu, Zhu, Chen, Sun, and Zhou}{Wei et~al\mbox{.}}{2021}]%
        {wei2021logreducer}
\bibfield{author}{\bibinfo{person}{Junyu Wei}, \bibinfo{person}{Guangyan Zhang}, \bibinfo{person}{Yang Wang}, \bibinfo{person}{Zhiwei Liu}, \bibinfo{person}{Zhanyang Zhu}, \bibinfo{person}{Junchao Chen}, \bibinfo{person}{Tingtao Sun}, {and} \bibinfo{person}{Qi Zhou}.} \bibinfo{year}{2021}\natexlab{}.
\newblock \showarticletitle{On the feasibility of parser-based log compression in Large-Scale cloud systems}. In \bibinfo{booktitle}{\emph{19th USENIX Conference on File and Storage Technologies (FAST 21)}}. \bibinfo{pages}{249--262}.
\newblock


\bibitem[\protect\citeauthoryear{Wikipedia}{Wikipedia}{2024}]%
        {Zstdwiki}
\bibfield{author}{\bibinfo{person}{Wikipedia}.} \bibinfo{year}{2024}\natexlab{}.
\newblock \bibinfo{title}{Zstandard}.
\newblock
\newblock
\urldef\tempurl%
\url{https://en.wikipedia.org/wiki/Zstd}
\showURL{%
\tempurl}


\bibitem[\protect\citeauthoryear{{Wikipedia contributors}}{{Wikipedia contributors}}{2023}]%
        {Deflate}
\bibfield{author}{\bibinfo{person}{{Wikipedia contributors}}.} \bibinfo{year}{2023}\natexlab{}.
\newblock \bibinfo{title}{Deflate --- {Wikipedia}{,} The Free Encyclopedia}.
\newblock
\newblock
\urldef\tempurl%
\url{https://en.wikipedia.org/w/index.php?title=Deflate&oldid=1148886022}
\showURL{%
\tempurl}


\bibitem[\protect\citeauthoryear{Yao, B.~de P{\'a}dua, Shang, Sporea, Toma, and Sajedi}{Yao et~al\mbox{.}}{2018}]%
        {yao2018log4perf}
\bibfield{author}{\bibinfo{person}{Kundi Yao}, \bibinfo{person}{Guilherme B.~de P{\'a}dua}, \bibinfo{person}{Weiyi Shang}, \bibinfo{person}{Steve Sporea}, \bibinfo{person}{Andrei Toma}, {and} \bibinfo{person}{Sarah Sajedi}.} \bibinfo{year}{2018}\natexlab{}.
\newblock \showarticletitle{Log4perf: Suggesting logging locations for web-based systems' performance monitoring}. In \bibinfo{booktitle}{\emph{Proceedings of the 2018 ACM/SPEC International Conference on Performance Engineering}}. \bibinfo{pages}{127--138}.
\newblock


\bibitem[\protect\citeauthoryear{Yu, Chen, Li, Weng, Zheng, Deng, and Zheng}{Yu et~al\mbox{.}}{2023}]%
        {yu2023logreducer}
\bibfield{author}{\bibinfo{person}{Guangba Yu}, \bibinfo{person}{Pengfei Chen}, \bibinfo{person}{Pairui Li}, \bibinfo{person}{Tianjun Weng}, \bibinfo{person}{Haibing Zheng}, \bibinfo{person}{Yuetang Deng}, {and} \bibinfo{person}{Zibin Zheng}.} \bibinfo{year}{2023}\natexlab{}.
\newblock \showarticletitle{Logreducer: Identify and reduce log hotspots in kernel on the fly}. In \bibinfo{booktitle}{\emph{2023 IEEE/ACM 45th International Conference on Software Engineering (ICSE)}}. IEEE, \bibinfo{pages}{1763--1775}.
\newblock


\bibitem[\protect\citeauthoryear{Zhang, Peng, Sha, Zhang, Fu, Wu, Lin, and Zhang}{Zhang et~al\mbox{.}}{2022}]%
        {zhang2022deeptralog}
\bibfield{author}{\bibinfo{person}{Chenxi Zhang}, \bibinfo{person}{Xin Peng}, \bibinfo{person}{Chaofeng Sha}, \bibinfo{person}{Ke Zhang}, \bibinfo{person}{Zhenqing Fu}, \bibinfo{person}{Xiya Wu}, \bibinfo{person}{Qingwei Lin}, {and} \bibinfo{person}{Dongmei Zhang}.} \bibinfo{year}{2022}\natexlab{}.
\newblock \showarticletitle{Deeptralog: Trace-log combined microservice anomaly detection through graph-based deep learning}. In \bibinfo{booktitle}{\emph{Proceedings of the 44th international conference on software engineering}}. \bibinfo{pages}{623--634}.
\newblock


\bibitem[\protect\citeauthoryear{Zhang, Shen, Yang, Meng, Xiao, Jia, Li, Sun, Zhang, and Lin}{Zhang et~al\mbox{.}}{2023}]%
        {zhang2023high}
\bibfield{author}{\bibinfo{person}{Jiujing Zhang}, \bibinfo{person}{Zhitao Shen}, \bibinfo{person}{Shiyu Yang}, \bibinfo{person}{Lingkai Meng}, \bibinfo{person}{Chuan Xiao}, \bibinfo{person}{Wei Jia}, \bibinfo{person}{Yue Li}, \bibinfo{person}{Qinhui Sun}, \bibinfo{person}{Wenjie Zhang}, {and} \bibinfo{person}{Xuemin Lin}.} \bibinfo{year}{2023}\natexlab{}.
\newblock \showarticletitle{High-Ratio Compression for Machine-Generated Data}.
\newblock \bibinfo{journal}{\emph{Proceedings of the ACM on Management of Data}} \bibinfo{volume}{1}, \bibinfo{number}{4} (\bibinfo{year}{2023}), \bibinfo{pages}{1--27}.
\newblock


\bibitem[\protect\citeauthoryear{Zheng, Gao, Wan, Yan, Hu, Liu, Gao, Zhou, and Jensen}{Zheng et~al\mbox{.}}{2023}]%
        {ZhengGWYHLG0J23}
\bibfield{author}{\bibinfo{person}{Bolong Zheng}, \bibinfo{person}{Yongyong Gao}, \bibinfo{person}{Jingyi Wan}, \bibinfo{person}{Lingsen Yan}, \bibinfo{person}{Long Hu}, \bibinfo{person}{Bo Liu}, \bibinfo{person}{Yunjun Gao}, \bibinfo{person}{Xiaofang Zhou}, {and} \bibinfo{person}{Christian~S. Jensen}.} \bibinfo{year}{2023}\natexlab{}.
\newblock \showarticletitle{DecLog: Decentralized Logging in Non-Volatile Memory for Time Series Database Systems}.
\newblock \bibinfo{journal}{\emph{Proc. {VLDB} Endow.}} \bibinfo{volume}{17}, \bibinfo{number}{1} (\bibinfo{year}{2023}), \bibinfo{pages}{1--14}.
\newblock


\bibitem[\protect\citeauthoryear{Zhou, Peng, Xie, Sun, Ji, Liu, Xiang, and He}{Zhou et~al\mbox{.}}{2019}]%
        {zhou2019latent}
\bibfield{author}{\bibinfo{person}{Xiang Zhou}, \bibinfo{person}{Xin Peng}, \bibinfo{person}{Tao Xie}, \bibinfo{person}{Jun Sun}, \bibinfo{person}{Chao Ji}, \bibinfo{person}{Dewei Liu}, \bibinfo{person}{Qilin Xiang}, {and} \bibinfo{person}{Chuan He}.} \bibinfo{year}{2019}\natexlab{}.
\newblock \showarticletitle{Latent error prediction and fault localization for microservice applications by learning from system trace logs}. In \bibinfo{booktitle}{\emph{Proceedings of the 2019 27th ACM joint meeting on European software engineering conference and symposium on the foundations of software engineering}}. \bibinfo{pages}{683--694}.
\newblock


\bibitem[\protect\citeauthoryear{Zhu, He, He, Liu, and Lyu}{Zhu et~al\mbox{.}}{2023}]%
        {Loghub}
\bibfield{author}{\bibinfo{person}{Jieming Zhu}, \bibinfo{person}{Shilin He}, \bibinfo{person}{Pinjia He}, \bibinfo{person}{Jinyang Liu}, {and} \bibinfo{person}{Michael~R. Lyu}.} \bibinfo{year}{2023}\natexlab{}.
\newblock \showarticletitle{Loghub: {A} Large Collection of System Log Datasets for AI-driven Log Analytics}. In \bibinfo{booktitle}{\emph{IEEE International Symposium on Software Reliability Engineering (ISSRE)}}.
\newblock


\bibitem[\protect\citeauthoryear{Ziv and Lempel}{Ziv and Lempel}{1977}]%
        {ziv1977lz77}
\bibfield{author}{\bibinfo{person}{Jacob Ziv} {and} \bibinfo{person}{Abraham Lempel}.} \bibinfo{year}{1977}\natexlab{}.
\newblock \showarticletitle{A universal algorithm for sequential data compression}.
\newblock \bibinfo{journal}{\emph{IEEE Transactions on information theory}} \bibinfo{volume}{23}, \bibinfo{number}{3} (\bibinfo{year}{1977}), \bibinfo{pages}{337--343}.
\newblock


\bibitem[\protect\citeauthoryear{Ziv and Lempel}{Ziv and Lempel}{1978}]%
        {ziv1978lz78}
\bibfield{author}{\bibinfo{person}{Jacob Ziv} {and} \bibinfo{person}{Abraham Lempel}.} \bibinfo{year}{1978}\natexlab{}.
\newblock \showarticletitle{Compression of individual sequences via variable-rate coding}.
\newblock \bibinfo{journal}{\emph{IEEE transactions on Information Theory}} \bibinfo{volume}{24}, \bibinfo{number}{5} (\bibinfo{year}{1978}), \bibinfo{pages}{530--536}.
\newblock


\end{thebibliography}

\end{document}